%% file: main.tex
\PassOptionsToPackage{table}{xcolor}
\documentclass[letterpaper]{article}
\usepackage{aaai2027}
\nocopyright

\usepackage[hyphens]{url}
\usepackage{graphicx}
\usepackage{natbib}
\usepackage{caption}
\usepackage{booktabs}
\usepackage{amsfonts}
\usepackage{amsmath}
\usepackage{amssymb}
\usepackage{array}
\usepackage{algorithm}
\usepackage{algorithmic}
\usepackage{overpic}
\usepackage{subcaption}
\usepackage{diagbox}
\usepackage{multirow}
\usepackage{colortbl}
\usepackage{wasysym}
\usepackage{xspace}
\usepackage{xcolor}
\usepackage{enumitem}
\usepackage{pifont}
\usepackage{siunitx}

\definecolor{darkgreen}{RGB}{0,160,0}
\definecolor{darkred}{RGB}{178,34,34}

\definecolor{mycolor1}{rgb}{1, 1, 1} % White
\definecolor{mycolor2}{rgb}{0.941, 0.939, 0.969} % Lightened purple
\definecolor{mycolor3}{rgb}{0.882, 0.876, 0.937} % Medium lightened purple
\definecolor{mycolor4}{rgb}{0.780, 0.783, 0.890} % Darker lightened purple

\providecommand{\autoref}[1]{\ref{#1}}

\newcommand{\etc}{\textit{etc.}\xspace}
\newcommand{\eg}{\textit{e.g.}\xspace}
\newcommand{\ie}{\textit{i.e.}\xspace}

\newcommand{\sysname}{\textsc{VIPatch}\xspace}

\title{\textit{Hiding in Plain Sight}: An Effective Physical Adversarial Patch Attack against Visual-Infrared Fused Face Detection}
\author{
Qiucheng Yu\textsuperscript{\rm 1},
Tao Ni\textsuperscript{\rm 2},
Yihe Zhou\textsuperscript{\rm 1},
Jiayimei Wang\textsuperscript{\rm 1},
Qingchuan Zhao\textsuperscript{\rm 1}
}

\affiliations{
\textsuperscript{\rm 1}City University of Hong Kong\\
\textsuperscript{\rm 2}King Abdullah University of Science and Technology
(KAUST)\\
}

\begin{document}
\maketitle

\begin{abstract}
% Deep learning-based visual-infrared fused face detection models are increasingly utilized across various applications but are susceptible to adversarial patch attacks. Most previous attacks have primarily targeted either the visual or infrared image alone in the digital domain, proving ineffective against visual-infrared fused face detection models in the physical world, and most of these existing methods are suspicious as their patched pattern significantly deviates from real-world patterns.
% In this paper, we introduce a novel physical adversarial patch attack, named \sysname (\underline{\textbf{V}}isual-\underline{\textbf{I}}nfrared \underline{\textbf{Patch}}), which produces inconspicuous, realistic, and natural-looking patches for facial images. Specifically, this is accomplished by creating a gradient color mask and applying a band-aid sticker in both the visual and infrared images, with joint optimization of these two elements, and the generated digital patches could also guide the creation of physical patches.
% Experimental results show that our method achieves competitive overall ASRs (over 90\%) in both digital and physical domains, and the patches created by our method are designed to attract minimal attention.
% \looseness=-1
% \looseness=-1
% {https://anonymous.4open.science/r/VIPatch}
Deep learning-based visual-infrared fused face detection models are increasingly deployed across a wide range of applications, yet they remain susceptible to adversarial patch attacks. Most prior attacks target either the visual or the infrared image alone in the digital domain, which renders them ineffective against fused models in the physical world. Moreover, many of these methods are readily noticeable, as their patch patterns deviate substantially from those seen in the real world. In this paper, we introduce \sysname (\underline{\textbf{V}}isual-\underline{\textbf{I}}nfrared \underline{\textbf{Patch}}), a novel physical adversarial patch attack that produces inconspicuous, realistic, and natural-looking patches for facial images. Specifically, \sysname crafts a gradient-color mask together with a band-aid sticker across both the visual and infrared images, and jointly optimizes these two elements; the resulting digital patches further guide the fabrication of their physical counterparts. Experimental results show that \sysname achieves competitive attack success rates (over 90\%) in both the digital and physical domains, while keeping the patches unobtrusive to human observers.

\looseness=-1
\end{abstract}

% \begin{links}
%     \link{Code}{https://anonymous.4open.science/r/VIPatch}
% \end{links}

\section{Introduction}
\label{Sec:Introduction}
% Face detection has been applied in various applications, including biometric identification~\cite{zulfiqar2019deep,ranjan2019fast} and security surveillance~\cite{yakovleva2023face,jian2010face}. 
% However, traditional visual face detection systems encounter performance degradation under challenging conditions~\cite{kumar2019face,singh2018techniques} such as low-light environments, adverse weather, and occlusions. 
% To overcome these limitations, visual-infrared fused face detection models~\cite{mohd2014facial,wang2007performance,ouyang2016survey,oh2017gabor} have emerged as a powerful alternative, combining visual and infrared data to ensure reliability across diverse scenarios. For example, during the COVID-19 pandemic outbreak period, temperature screening systems were utilized in high-traffic areas such as airports to detect fevers efficiently and quickly. To minimize false alarms and interference from other heat sources (\eg, mobile phones, power banks, and pets), these systems~\cite{guidesense200qt,covid2020,dlinkthermal} typically adopt deep learning-based visual-infrared fused face detection models to identify facial regions by detecting the face from both visual and infrared images captured by infrared cameras,
% and then detect abnormal temperatures within those regions.
% \looseness=-1

\begin{figure}[t]
        \centering
        \includegraphics[width=\linewidth]{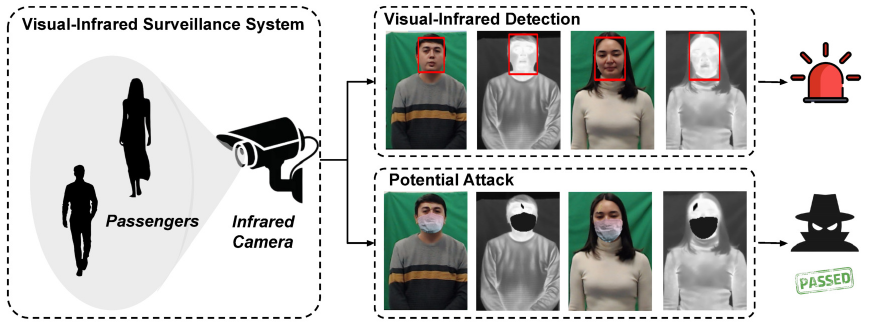}
        \vspace{-0.2in}
        \caption{An adversary spoofs and bypasses the visual-infrared fused detection system with \sysname.}
        \vspace{-0.2in}
        \label{thermal camera}
\end{figure}

Face detection has been applied in various applications, including biometric identification~\cite{zulfiqar2019deep} and security surveillance~\cite{yakovleva2023face}. However, traditional visual face detection systems encounter performance degradation under challenging conditions~\cite{kumar2019face,singh2018techniques} such as low-light environments, adverse weather, and occlusions. To overcome these limitations, visual-infrared fused face detection models~\cite{mohd2014facial,wang2007performance,ouyang2016survey,oh2017gabor} have emerged as a powerful alternative that combines visual and infrared data to ensure reliability across diverse scenarios. For example, during the COVID-19 pandemic, temperature screening systems were deployed in high-traffic areas such as airports to detect fevers quickly and efficiently. To minimize false alarms and interference from other heat sources (\eg, mobile phones, power banks, and pets)~\cite{ni2023eavesdropping, ni2023uncovering, ni2023exploiting, ni2023xporter}, these systems~\cite{guidesense200qt,covid2020,dlinkthermal} typically adopt deep learning-based visual-infrared fused face detection models, which locate facial regions in both visual and infrared images captured by infrared cameras and then detect abnormal temperatures within those regions. Unfortunately, since such models are fundamentally deep-learning-based, they inherit the vulnerability of deep networks to adversarial attacks~\cite{doan2022tnt,qian2020visually, yuan2024itpatch, yuan2025no, wang2026adversarial}. In their application scenario, this suggests that the fused face detection model is highly likely susceptible to adversarial patch attacks, which operate by introducing perceptible perturbations in the physical world~\cite{ni2023recovering, ni2024non, ni2024rehsense, sun2025spacesched, ni2024sensor, ni2025characterizing, ni2025good, wu2025ringbyte, ni2026vr, zhou2025survey, wang2025contemporary, sun2024rf, yuan2025fighost, yuan2026fluorescent, meng2024ava, ni2021simple, zhao2022periscope, guo2024puridefense, zhang2025hidden}. Following this hypothesis, we can intuitively expect that an adversary could bypass the detection simply by wearing a mask displaying particular geometric figures, as shown in Fig.~\autoref{thermal camera}.
\looseness=-1

% \begin{figure*}[t]
%     \minipage{0.49\textwidth}%
%         \centering
%         \includegraphics[width=\linewidth]{pictures/methodology/illustration_ifpatch.pdf}
%         \caption{An adversary spoofs and bypasses the visual-infrared fused detection system by wearing a mask or placing a band-aid on the body.}
%         \label{thermal camera}
%     \endminipage\hfill
%     \minipage{0.49\textwidth}%
%         \centering
%         \includegraphics[width=\linewidth]{pictures/methodology/illustrative_example.pdf}
%         \caption{Physical adversarial attacks with visual patches in prior works~\cite{nguyen2020adversarial,yang2020design,komkov2021advhat,yin2021adv,pautov2019adversarial,sharif2016accessorize} and our visual-infrared fused patches in \sysname.}
%         \label{physical attack examples}
%     \endminipage
% \end{figure*}

% Unfortunately, visual-infrared fused face detection models are fundamentally deep-learning-based models that are inherently susceptible to adversarial attacks~\cite{jan2019connecting,doan2022tnt,qian2020visually}.
% As such, particularly in its application scenario, this fused face detection model is highly likely vulnerable to adversarial patch attacks, which directly launch attacks by creating obvious perturbations in the physical world.
% Following this hypothesis, we can intuitively expect that an adversary could bypass the detection by simply wearing a mask displaying particular geometric figures, as shown in Fig.~\autoref{thermal camera}.

However, \textit{no previous study has explored the presence of this adversarial attack in real-world applications atop this fused face detection model, and no prior works investigate such a vulnerability.}
In particular, \textit{($i$)} previous patch attacks ~\cite{Adv-Patch,Adv-Cloak,Adv-Texture,sharif2016accessorize,komkov2021advhat,adv-mask,AIP,zhu2022infrared} targeted either the visual or the infrared images by optimizing their patches based on the gradient of the model, and they thus \textit{cannot} apply to the visual infrared fused model directly.
Moreover, \textit{($ii$)} some other works~\cite{rs-patch,chen2020hopskipjumpattack,li2020qeba,hardbeat,RL-patch} were not evaluated in the physical domain but entirely simulated the attack in the digital domain by perturbing the pixels, which makes them unable to unveil the real-world impacts.
In addition, \textit{($iii$)} some of these meticulously crafted patches' patterns significantly deviate from real-world patterns, making them appear suspicious and easily detectable in real-world scenarios.
Therefore, \textit{there is a huge gap and urgent need in the research on the robustness of visual infrared fused face detection models in real-world applications.}
\looseness=-1

% To fill this gap, we plan to explore the feasibility of applying an adversarial patch attack to bypass real-world face detection systems deploying visual-infrared fused models.
% In particular, this attack should be
% \textit{(i) a joint attack} that requires the patch to be jointly optimized in both the visual and the infrared images, 
% \textit{(ii) a physical attack} where the patch optimization should consider the physical constraints
% to enable the patch attack in the physical world,
% and \textit{(iii) a natural and stealthy attack} in which the patch should be designed to be common in real scenarios to avoid attracting attention.
% As outlined in Table~\autoref{tab:comparision}, \textit{these three requirements make our proposed attack novel and advance over previous studies.}
% \looseness=-1

To fill this gap, we design and implement \sysname, a novel physical adversarial patch attack that bypasses real-world face detection systems deploying visual-infrared fused models. The core idea is to craft the patch from a gradient-color mask and a band-aid sticker, both of which are commonplace in the real world, so that the attack remains inconspicuous. Building on this idea, our design fulfills three requirements. To begin with, as a \textit{physical attack}, we fabricate an actual mask and band-aid and account for real-world transformations to bridge the gap between the digital and physical domains, after which we harmonize the embedded images with the surrounding lighting and color. In terms of \textit{stealthiness}, unlike prior physical attacks~\cite{nguyen2020adversarial,yang2020design,komkov2021advhat,yin2021adv,pautov2019adversarial,sharif2016accessorize} that rely on unrealistic patches, we optimize the mask's color under color-harmonization constraints to obtain a visually pleasing gradient, and we further confine the band-aid to a valid area that avoids covering key facial features such as the eyes, thereby reducing the chance of attracting attention. Finally, to launch a \textit{joint attack} across both the visual and infrared images, we develop and refine the mask and band-aid for each modality, which prevents the aforementioned conflicts and enhances effectiveness. As outlined in Table~\autoref{tab:comparision}, these three properties make \sysname novel and advance over previous studies.
\looseness=-1

% \input{tables/compare_methods3}

% It is by no means trivial to realize this adversarial physical patch attack. 
% The first challenge resides in the \textit{stealthiness of the physical attack} because previous methods did not optimize patches jointly across visual and infrared images. 
% As such, merely applying all patches in their respective images to the face results in excessive patches, which become overly conspicuous and attract undue attention. 
% The second uncertainty is \textit{effectiveness of the joint attack}.
% An intuitive solution is to combine existing patches from the visual and infrared images directly. However, placing patches on both the visual and infrared images could lead to conflicts, such as overlapping patches at the same location. 
% That is, placing a visual patch on top of an infrared patch may result in unintended alterations in the infrared image, whereas positioning an infrared patch over a visual patch may prevent the visual patch from being fully detected by the visual camera.
% \looseness=-1

\begin{table}[t]
\centering
\scriptsize
\setlength{\tabcolsep}{1pt}
\renewcommand{\arraystretch}{0.8}
\caption{Comparison with prior adversarial patch attacks. $\square$: White-box, $\blacksquare$: Black-box, ASR: Attack success rate.}
\label{tab:comparision}
\vspace{-0.05in}
\resizebox{\linewidth}{!}{
\begin{tabular}{lcccccccccccc}
%{\hsize}{@{}@{\extracolsep{\fill}}cccccccccccccc@{}}
\toprule
\textbf{Target Model} & \textbf{Physically}&\textbf{Stealthiness}&\textbf{Jointly}&\textbf{$\square$~ASR}&\textbf{$\blacksquare$~ASR}&\textbf{Physical}  \\
\midrule\midrule

%\hline 

\multicolumn{1}{l|}{RL-Patch~\cite{RL-patch}}&\multicolumn{1}{c|}{ \textcolor{darkred}{\ding{56}}}&\multicolumn{1}{c|}{ \textcolor{darkred}{\ding{56}}}&\multicolumn{1}{c|}{ \textcolor{darkred}{\ding{56}}}& \multicolumn{1}{c|}{--} & \multicolumn{1}{c|}{72.68} & \multicolumn{1}{c}{--} \\

\multicolumn{1}{l|}{HardBeat~\cite{hardbeat}}&\multicolumn{1}{c|}{ \textcolor{darkred}{\ding{56}}} &\multicolumn{1}{c|}{ \textcolor{darkred}{\ding{56}}} &\multicolumn{1}{c|}{\textcolor{darkred}{\ding{56}}} & \multicolumn{1}{c|}{--} & \multicolumn{1}{c|}{86.95} & \multicolumn{1}{c}{--}   \\

\multicolumn{1}{l|}{RS-Patch~\cite{rs-patch}}&\multicolumn{1}{c|}{ \textcolor{darkred}{\ding{56}}} &\multicolumn{1}{c|}{ \textcolor{darkred}{\ding{56}}} &\multicolumn{1}{c|}{ \textcolor{darkred}{\ding{56}}} & \multicolumn{1}{c|}{--} & \multicolumn{1}{c|}{93.13} & \multicolumn{1}{c}{--}   \\ 

\multicolumn{1}{l|}{AIP~\cite{AIP}}&\multicolumn{1}{c|}{\textcolor{darkgreen}{\ding{52}}}&\multicolumn{1}{c|}{\textcolor{darkred}{\ding{56}}}&\multicolumn{1}{c|}{\textcolor{darkred}{\ding{56}}} & \multicolumn{1}{c|}{92.36} & \multicolumn{1}{c|}{64.12} & \multicolumn{1}{c}{68.05}  \\

\multicolumn{1}{l|}{HCB~\cite{HCB}}&\multicolumn{1}{c|}{ \textcolor{darkgreen}{\ding{52}}} &\multicolumn{1}{c|}{ \textcolor{darkgreen}{\ding{52}}} &\multicolumn{1}{c|}{ \textcolor{darkred}{\ding{56}}} & \multicolumn{1}{c|}{--} & \multicolumn{1}{c|}{92.33} & \multicolumn{1}{c}{55.63}   \\

\multicolumn{1}{l|}{AdvIB~\cite{AdvIB}}&\multicolumn{1}{c|}{ \textcolor{darkgreen}{\ding{52}}} &\multicolumn{1}{c|}{ \textcolor{darkgreen}{\ding{52}}} &\multicolumn{1}{c|}{ \textcolor{darkred}{\ding{56}}} & \multicolumn{1}{c|}{--} & \multicolumn{1}{c|}{91.39} & \multicolumn{1}{c}{86.81}   \\

\multicolumn{1}{l|}{Adv-Patch~\cite{Adv-Patch}}&\multicolumn{1}{c|}{ \textcolor{darkgreen}{\ding{52}}} &\multicolumn{1}{c|}{ \textcolor{darkred}{\ding{56}}} &\multicolumn{1}{c|}{ \textcolor{darkred}{\ding{56}}} & \multicolumn{1}{c|}{95.26} & \multicolumn{1}{c|}{60.37} & \multicolumn{1}{c}{--}   \\

\multicolumn{1}{l|}{Adv-Cloak~\cite{Adv-Cloak}}&\multicolumn{1}{c|}{ \textcolor{darkgreen}{\ding{52}}} &\multicolumn{1}{c|}{ \textcolor{darkred}{\ding{56}}} &\multicolumn{1}{c|} {\textcolor{darkred}{\ding{56}}} & \multicolumn{1}{c|}{63.39} & \multicolumn{1}{c|}{49.21} & \multicolumn{1}{c}{50.00}   \\

\multicolumn{1}{l|}{Adv-Texture~\cite{Adv-Texture}}&\multicolumn{1}{c|}{ \textcolor{darkgreen}{\ding{52}}} &\multicolumn{1}{c|}{ \textcolor{darkred}{\ding{56}}} &\multicolumn{1}{c|}{ \textcolor{darkred}{\ding{56}}} & \multicolumn{1}{c|}{65.57} & \multicolumn{1}{c|}{51.36} & \multicolumn{1}{c}{64.10}   \\

\multicolumn{1}{l|}{Adv-Mask~\cite{adv-mask}}&\multicolumn{1}{c|}{ \textcolor{darkgreen}{\ding{52}}} &\multicolumn{1}{c|}{ \textcolor{darkred}{\ding{56}}} &\multicolumn{1}{c|}{ \textcolor{darkred}{\ding{56}}} & \multicolumn{1}{c|}{62.80} & \multicolumn{1}{c|}{67.07} & \multicolumn{1}{c}{73.30}   \\

\multicolumn{1}{l|}{Adv-Sticker~\cite{adv-sticker}}&\multicolumn{1}{c|}{ \textcolor{darkgreen}{\ding{52}}} &\multicolumn{1}{c|}{ \textcolor{darkgreen}{\ding{52}}} &\multicolumn{1}{c|}{ \textcolor{darkred}{\ding{56}}} & \multicolumn{1}{c|}{25.27} & \multicolumn{1}{c|}{40.57} & \multicolumn{1}{c}{--}   \\

% \multicolumn{1}{l|}{RL-Patch~\cite{RL-patch}}&\multicolumn{1}{c|}{ \textcolor{darkred}{\ding{56}}}&\multicolumn{1}{c|}{ \textcolor{darkred}{\ding{56}}}&\multicolumn{1}{c}{ \textcolor{darkred}{\ding{56}}}&  \\

% \multicolumn{1}{l|}{HardBeat~\cite{hardbeat}}&\multicolumn{1}{c|}{ \textcolor{darkred}{\ding{56}}} &\multicolumn{1}{c|}{ \textcolor{darkred}{\ding{56}}} &\multicolumn{1}{c}{\textcolor{darkred}{\ding{56}}}   \\

% \multicolumn{1}{l|}{RS-Patch~\cite{rs-patch}}&\multicolumn{1}{c|}{ \textcolor{darkred}{\ding{56}}} &\multicolumn{1}{c|}{ \textcolor{darkred}{\ding{56}}} &\multicolumn{1}{c}{ \textcolor{darkred}{\ding{56}}}   \\ 

\multicolumn{1}{l|}{UVHat~\cite{yuanomni}}&\multicolumn{1}{c|}{ \textcolor{darkgreen}{\ding{52}}} &\multicolumn{1}{c|}{ \textcolor{darkgreen}{\ding{52}}} &\multicolumn{1}{c|}{ \textcolor{darkred}{\ding{56}}} & \multicolumn{1}{c|}{--} & \multicolumn{1}{c|}{--} & \multicolumn{1}{c}{77.13}   \\ 

\multicolumn{1}{l|}{ProjAttacker~\cite{liu2025projattacker}}&\multicolumn{1}{c|}{  \textcolor{darkgreen}{\ding{52}}} &\multicolumn{1}{c|}{ \textcolor{darkgreen}{\ding{52}}} &\multicolumn{1}{c|}{ \textcolor{darkred}{\ding{56}}} & \multicolumn{1}{c|}{--} & \multicolumn{1}{c|}{83.80} & \multicolumn{1}{c}{63.33}   \\ 

\multicolumn{1}{l|}{ARA~\cite{zhang2024adversarial}}&\multicolumn{1}{c|}{  \textcolor{darkgreen}{\ding{52}}} &\multicolumn{1}{c|}{ \textcolor{darkgreen}{\ding{52}}} &\multicolumn{1}{c|}{ \textcolor{darkred}{\ding{56}}} & \multicolumn{1}{c|}{96.33} & \multicolumn{1}{c|}{85.98} & \multicolumn{1}{c}{--}   \\ 

\multicolumn{1}{l|}{Face3DAdv~\cite{yang2025face3dadv}}&\multicolumn{1}{c|}{  \textcolor{darkgreen}{\ding{52}}} &\multicolumn{1}{c|}{ \textcolor{darkgreen}{\ding{52}}} &\multicolumn{1}{c|}{ \textcolor{darkred}{\ding{56}}} & \multicolumn{1}{c|}{93.76} & \multicolumn{1}{c|}{63.36} & \multicolumn{1}{c}{--}   \\ 

\multicolumn{1}{l|}{Agile~\cite{wang2024invisible}}&\multicolumn{1}{c|}{  \textcolor{darkgreen}{\ding{52}}} &\multicolumn{1}{c|}{ \textcolor{darkgreen}{\ding{52}}} &\multicolumn{1}{c|}{ \textcolor{darkred}{\ding{56}}} & \multicolumn{1}{c|}{--} & \multicolumn{1}{c|}{--} & \multicolumn{1}{c}{84.20}   \\ 

\midrule

% \multicolumn{1}{l|}{\textbf{VIPatch(ours)}}&\multicolumn{1}{c}{ \textcolor{darkgreen}{\ding{52}}} &\multicolumn{1}{c}{ \textcolor{darkgreen}{\ding{52}}} &\multicolumn{1}{c}{ \textcolor{darkgreen}{\ding{52}}}   \\
\multicolumn{1}{l|}{\textbf{Our Attack}}&\multicolumn{1}{c|}{ \textcolor{darkgreen}{\ding{52}}} &\multicolumn{1}{c|}{ \textcolor{darkgreen}{\ding{52}}} &\multicolumn{1}{c|}{ \textcolor{darkgreen}{\ding{52}}} & \multicolumn{1}{c|}{--} & \multicolumn{1}{c|}{94.59} & \multicolumn{1}{c}{93.83}   \\

%\hline
%\hline
%\multicolumn{1}{c|}{\textbf{Difficult Defense}}& \multicolumn{1}{c}{\textbf{--}}&\multicolumn{1}{c}{\textbf{--}}&\multicolumn{1}{c}{\textbf{--}}&\multicolumn{1}{c}{\textcolor{darkred}\textcolor{darkred}{\ding{56}}} &\multicolumn{1}{c}{\textcolor{darkred}\textcolor{darkred}{\ding{56}}} &\multicolumn{1}{c}{\textcolor{darkred}\textcolor{darkred}{\ding{56}}} &\multicolumn{1}{c}{\textcolor{darkgreen}\textcolor{darkgreen}{\ding{52}}} &\multicolumn{1}{c}{\textcolor{darkgreen}\textcolor{darkgreen}{\ding{52}}} &\multicolumn{1}{c}{\textcolor{darkred}\textcolor{darkred}{\ding{56}}} &\multicolumn{1}{c}{\textcolor{darkred}\textcolor{darkred}{\ding{56}}} &\multicolumn{1}{c}{\textcolor{darkred}\textcolor{darkred}{\ding{56}}}&\multicolumn{1}{c}{\textcolor{darkgreen}\textcolor{darkgreen}{\ding{52}}}  \\

%\hline
\bottomrule
\end{tabular}
}
% \caption{Comparison between other methods and \sysname.}
% \caption{Comparison between other SOTA methods and ours.}
\vspace{-0.2in}
% \label{comparision}
\end{table}

% In this paper, we designed and implemented \sysname, a novel physical adversarial patch attack that leverages a common mask with gradient color and band-aid sticker to generate a physical patch 
% to bypass the visual-infrared face detection model.
% In particular, to launch the \textit{physical attack}, we first create an actual mask and band-aid, taking into account real-world transformations to bridge the gap between digital and physical domains. We then adjust the embedded masked images to ensure the integrated images are harmonized with the lighting conditions and color differences, achieving a harmonic visual effect. 
% Then, to enhance \textit{stealthiness}, unlike previous physical attacks~\cite{nguyen2020adversarial,yang2020design,komkov2021advhat,yin2021adv,pautov2019adversarial,sharif2016accessorize} that exploit unrealistic visual patches, we utilize a real mask and band-aid, both commonplace in the real world, thus reducing the chances of attracting attention. Also, we optimize the mask's color within color harmonization constraints to create a gradient color mask that is visually pleasing to observers. In addition, we designate a valid area for the band-aid to ensure it does not cover key facial features like eyes.
% Finally, to launch a \textit{joint attack} across both the visual and infrared images, we develop and refine the mask and band-aid for each image to prevent the previously mentioned conflicts and enhance effectiveness.
% \looseness=-1

\paragraph{Contributions.} Our contributions are listed as follows
% \footnote{Anonymous artifact of code and data are available at: \url{https://anonymous.4open.science/r/VIPatch}}
:

\begin{itemize}[label=\textbullet]
    \item \textbf{Novel Physical and Joint Adversarial Attack:} 
    % To the best of our knowledge, we are the first to 
    We propose the first physical adversarial patch against visual-infrared fused face detection models by jointly optimizing adversarial patches in both visual and infrared images.\looseness=-1
    \item \textbf{Practical and Stealthy Adversarial Patch:} We design \sysname that utilizes practical masks and band-aid stickers to generate adversarial patches on both facial and infrared images by maintaining high attack success rates while raising minimal suspicions.
    \item \textbf{Comprehensive Evaluation:} We comprehensively evaluate \sysname in both digital and physical domains. The results demonstrate that \sysname outperforms several baselines and be resilient to environmental impacts.
\end{itemize}

\section{Threat Model}
\label{sec:attacking_scenarios}

% \paragraph{Visual-infrared Fused Face Detection} Visual-infrared fused face detection is a crucial technique that not only identifies and locates human faces in images or videos, but also generates heatmaps and detects body temperatures, which has been adopted in a range of applications, including recognizing facial attributes and identities. 
% For example, during the COVID-19 pandemic, visual-infrared fusion face detection systems became instrumental in efficient temperature screening at customs checkpoints, hospitals, airports, train stations, and other densely populated areas, as shown in \autoref{thermal camera}.
% \looseness=-1

\paragraph{Attack Scenarios.} We consider a common attack scenario in which the attacker tries to cause the failure of visual-infrared fused face detection systems used in facilities such as customs temperature monitoring~\cite{guidesense200qt,covid2020,dlinkthermal} or anti-theft surveillance~\cite{Surveillancethermal1, Surveillancethermal2} by placing adversarial patches
on the human face.
\looseness=-1

\paragraph{Attacker's Goal.} 
% In this study, an attacker tries to bypass a visual-infrared system by wearing an adversarial patch (\eg, mask, band-aid sticker) to result in false detection and spoof the visual-infrared fused face detection systems. 
To attack the visual-infrared fused face detection models, face detectors must be targeted in both visible and infrared images simultaneously, making face detection ineffective. 
In practice, this attack can be conceptualized as an optimization problem, with the patch patterns and their placements acting as variables to be manipulated. The optimization problem is:
% \begin{equation}
% \small
% \left\{
%      \begin{array}{lr}
%      \mathop{\arg\min}\limits_{\sigma_v}f_v( M\odot \sigma_v+(1-M)\odot X_v),\\
     
%      \mathop{\arg\min}\limits_{\sigma_i}f_i( M\odot \sigma_i+(1-M)\odot X_i), 
%      \end{array}
% \right.
% \end{equation}

\begin{equation}
\small
\left\{
\begin{array}{l}
\displaystyle \arg\min_{\sigma_v}\; f_v\!\big(M\odot \sigma_v+(1-M)\odot X_v\big),\\[2pt]
\displaystyle \arg\min_{\sigma_i}\; f_i\!\big(M\odot \sigma_i+(1-M)\odot X_i\big).
\end{array}
\right.
\end{equation}
where $\sigma_v$, $X_v$ represent the adversarial patch and image in the visual aspect, $\sigma_i$ and $X_i$ denote the adversarial patch and image in the infrared aspect, and $M$ represents the patch mask applied to both the visual and infrared aspects, respectively. 
The attacker's main goal is to optimize both the visual and infrared patches to diminish object confidence in face detection models in visual-infrared fused detection systems.

\paragraph{Attacker's Capability.} In this study, we consider the black-box attack setting. The attacker has no prior knowledge of the target visual-infrared face detection models, including their weight parameters, hyper-parameters, the training process, and the original training data, \etc
This means the attacker cannot observe the models' behavior and gradients.
Meanwhile, the only feedback the attacker can obtain from the target models is the output confidence of a given input, \ie, a pair of visual and infrared images.
% \looseness=-1
%
%
% Specifically, we assume that the attacker cannot access the original training data of the target models and can only exploit his/her own dataset to create the adversarial patches.
% In practice, the attacker can query the target models to obtain the output probabilities repeatedly, and we consider a budget for the number of queries to carry out each attack that is less than $10,000$, due to the constraints of the computational resources that the attacker can obtain in real-world scenarios, following the research line~\cite{chen2017zoo,rs-patch,guo2019simple, hardbeat}.
In practice, the attacker can query the target models to obtain the output probabilities repeatedly, and we consider a budget for the number of queries to carry out each attack that is less than 10,000, due to the constraints of the computational resources that the attacker can obtain in real-world scenarios, following the research line~\cite{chen2017zoo,rs-patch,guo2019simple, hardbeat}.
\looseness=-1

\begin{figure*}[t]
    \centering
    \includegraphics[width=\linewidth]{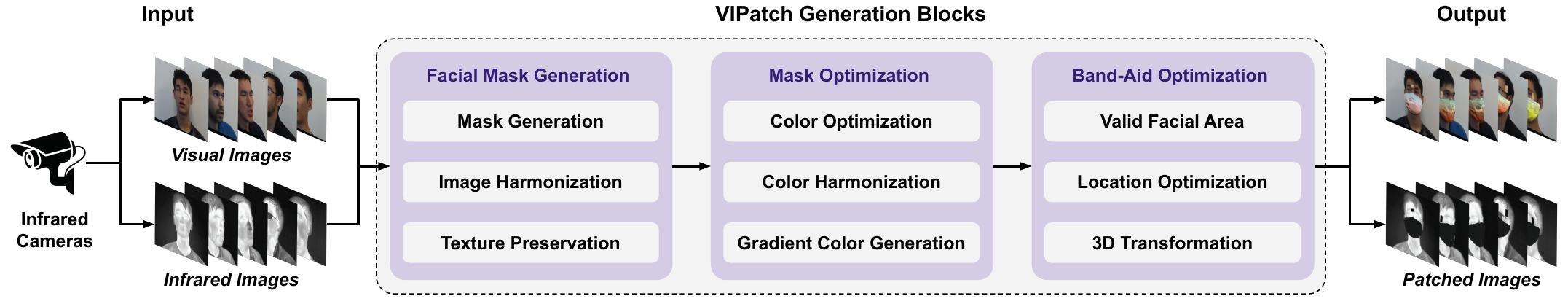}
    \vspace{-0.2in}
    \caption{Overview of \sysname.}
    \vspace{-0.15in}
    \label{fig:system_overview}
\end{figure*}

\section{\sysname Design}
\label{sec:system_design}

Fig.~\ref{fig:system_overview} depicts the overview of \sysname, which involves a black-box universal adversarial patch attack against visual-infrared fused face detection systems.
% \sysname strategically generates and applies a mask-wise patch and a band-aid sticker on both the visual and infrared images to deceive both the face detector and infrared detector in cameras of infrared surveillance systems.
\looseness=-1

\subsection{Facial Mask Generation}
\label{subsec:Facial Mask Generation}

\paragraph{Mask Generation.} 
% In previous black-box patch attacks~\cite{rs-patch,RL-patch,hardbeat}, the optimization of adversarial patches involved two distinct phases. Initially, the patch's content was fixed while its position was optimized. Subsequently, the content was further refined while maintaining the patch's position. Each of these phases incurred significant computational costs, highlighting the need to reduce the search space in optimization efforts.
% We adopted a strategy of using a fixed mask-wise patch. This approach eliminates the need to search for an optimal patch location. 
% Given that masks are commonly adorned with various prints, employing a mask-wise patch does not draw undue attention. Furthermore, mask patches, as opposed to traditional square patches, allow for the modification of a larger number of pixels in a less conspicuous manner, thus enhancing the attack's effectiveness.

% Previous black-box patch attacks~\cite{rs-patch,RL-patch,hardbeat} optimize adversarial patches in two phases: fixing the patch content while searching for its position, then refining the content at that fixed position. Both phases are computationally expensive, motivating the need to reduce the search space. To this end, we adopt a fixed mask-wise patch, which eliminates the search for an optimal location altogether. Since masks are commonly printed with various patterns, a mask-wise patch attracts little attention; moreover, compared with traditional square patches, it modifies far more pixels in a less conspicuous manner, thereby enhancing effectiveness.
Unlike previous black-box patch attacks~\cite{rs-patch,RL-patch,hardbeat} that separately search for the patch's position and content at significant computational cost, we adopt a fixed mask-wise patch, which eliminates the location search entirely. Since masks are commonly printed with various patterns, such a patch attracts little attention; moreover, compared with traditional square patches, it modifies far more pixels in a less conspicuous manner, thereby enhancing effectiveness.
\looseness=-1

\paragraph{Image Harmonization.} To minimize the disparity between the digital and physical domains, we employ actual masks rather than mask patterns and utilize UV mapping to apply them to the human face. UV mapping involves the projection of a 3D model's surface onto a 2D image for texture mapping, thereby preserving the 3D information of the images. In our approach, the facial landmarks are extracted using a face recognition method, and a normal mask is aligned with the face accordingly.
Then, the original face image and mask image are transformed into UV space using the 3D face reconstruction method. Subsequently, within the UV space, the face image and mask image are merged by rendering the UV texture map, and the complete face image with the applied mask is reconstructed.
\looseness=-1

However, we find that merely applying a uniform mask pattern on the embedded image yields inharmonious results. To address this issue, post-processing in the form of image harmonization is necessary. Hence, we employ Reinhard’s algorithm~\cite{reinhard2001color} to adjust the foreground color to match the background and the pre-trained neural network to further harmonize the facial mask, rendering it more natural and realistic in appearance.

\begin{figure}[t]
        \centering
        \begin{subfigure}[b]{0.49\linewidth}
             \centering
             \includegraphics[width=\linewidth]{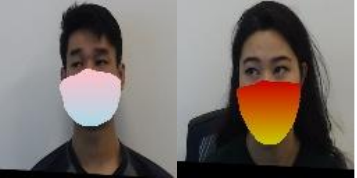}
             % \vspace{-0.15in}
             \caption{w/o Texture Preservation.}
             \label{fig:without_hsv}
        \end{subfigure}
        \begin{subfigure}[b]{0.49\linewidth}
             \centering
             \includegraphics[width=\linewidth]{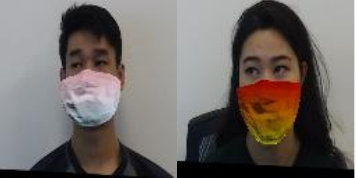}
             % \vspace{-0.15in}
             \caption{w/ Texture Preservation}
             \label{fig:with_hsv}
        \end{subfigure}
        \vspace{-0.1in}
        \caption{Patch images versus Texture Preservation.}
         \vspace{-0.15in}
        \label{fig:texture_preservation_patched_images}
\end{figure}

\paragraph{Texture Preservation.} Subsequently, we conduct texture preservation, which plays a vital role in minimizing the disparity between the digital and physical domains during the subsequent optimization steps. Fig.~\ref{fig:texture_preservation_patched_images} provides a visual comparison of the patched images with and without texture preservation, clearly demonstrating the unrealistic appearance of the image lacking texture preservation. To achieve texture preservation, we convert the original masked image into the HSV color space. In the HSV color space, the $V$ value represents the pixel brightness. Preserving the unaltered brightness of each pixel is of utmost importance to maintain the authentic texture of the mask. Consequently, only the $H$ and $S$ values undergo modification during the subsequent color optimization process.
Importantly, our method is designed for visual-infrared fused face detection models, necessitating joint optimization of the visual-infrared image pair. As a result, we incorporate a mask-wise cold block into the corresponding infrared image to account for the thermal insulation provided by the mask.
\looseness=-1

\subsection{Mask Optimization}
\label{subsec:Mask Optimization}
\paragraph{Color Optimization.} To minimize suspicion, we propose utilizing gradient color masks, which solely modify the mask color and avoid excessive alterations in other aspects. This approach is chosen for the availability and common use of gradient masks in daily life, avoiding drawing attention or arousing curiosity. 
The color combination is optimized based on the Differential Evolution (DE) algorithm ~\cite{storn1997differential}, which is efficient in searching for the optimum in the search space. The DE algorithm comprises four key steps: initialization, mutation, crossover, and selection. Each of these steps is comprehensively described as follows.
\looseness=-1

% \begin{figure*}[t]
%     \minipage{0.45\textwidth}%
%         \centering
%         \begin{subfigure}[b]{0.49\linewidth}
%              \centering
%              \includegraphics[width=\linewidth]{pictures/methodology/hsv_space_conversion_before.pdf}
%              % \vspace{-0.15in}
%              \caption{w/o Texture Preservation.}
%              \label{fig:without_hsv}
%         \end{subfigure}
%         \begin{subfigure}[b]{0.49\linewidth}
%              \centering
%              \includegraphics[width=\linewidth]{pictures/methodology/hsv_space_conversion_after.pdf}
%              % \vspace{-0.15in}
%              \caption{w/ Texture Preservation}
%              \label{fig:with_hsv}
%         \end{subfigure}
%         % \vspace{-0.1in}
%         \caption{Patch images versus Texture Preservation.}
%          % \vspace{-0.15in}
%         \label{fig:texture_preservation_patched_images}
%     \endminipage\hfill
%     \minipage{0.45\textwidth}%
%         \centering
%         \includegraphics[width=\linewidth]{pictures/methodology/color harmonization.png}
%         % \vspace{-0.23in}
%         \caption{Harmonic templates.}
%         % \vspace{-0.35in}
%         \label{color harmonization}
%     \endminipage
% \end{figure*}

\noindent\textit{(i) Initialization:} In initialization, we randomly generate $n$ color combinations, and $\theta$ and $n$ indicate one color combination and population size:
\begin{equation}
    \label{Initialization}
\small
    Blocks = {\theta_1, \theta_2,...,\theta_n}.
\end{equation}
% where $\theta$ and $n$ indicate one color combination and population size.
% \looseness=-1

\noindent\textit{(ii) Mutation:} The second step is the mutation. For every individual $\theta$ within the population, the mutated individual is obtained by adding it to the vector difference between two randomly selected individuals from the population, which can be formulated as:
\begin{equation}
\label{Mutation}
\small
    \theta_{r1} = Clipping(\theta_{r1} + \gamma_m* (\theta_{r2} - \theta_{r3})),
\end{equation}
where $r1,r2,r3 \in [1,2,...n], r1 \neq r2\neq r3$ and $Clipping$ mean that the parameters outside the constraint range are randomly generated within the range, and $\gamma_m$ indicates the mutation rate.
\looseness=-1

\noindent\textit{(iii) Crossover:} The next important step is the crossover. For each individual, a crossover operation is performed between the individual and its corresponding offspring mutation vector. More specifically, for each component, the offspring mutation vector is selected with a certain probability of replacing the original vector, generating a test individual. This process can be formulated as:
% \begin{equation}
% \label{Crossover}
% \small
% \theta^{j}_{r_1}=\left\{
%          \begin{array}{lr}
%          \theta^{p,j}_{r_1} & rand[0,1] \leq \gamma_c,\\
         
%          \theta^{j}_{r_1} & otherwise,
%          \end{array}
% \right.
% \end{equation}
\begin{equation}
\label{Crossover}
\small
\theta^{j,\text{new}}_{r_1}=\left\{
\begin{array}{ll}
\theta^{p,j}_{r_1}, & \mathrm{rand}(0,1)\le \gamma_c,\\
\theta^{j,\text{old}}_{r_1}, & \text{otherwise}.
\end{array}
\right.
\end{equation}
where $\theta^{j}_{r_1}$ represents the $j-th$ vector of the individual $\theta_{r_1}$ after mutation. $\theta^{p,j}_{r_1}$ is the $j-th$ vector of the individual before the mutation. $rand[0,1]$ represents a number randomly selected from 0 to 1, indicating random probability, and $\gamma_c$ represents the crossover rate. If the probability is not greater than $\gamma_c$, the $j-th$ vector of the individual $\theta_{r_1}$ remains unchanged. However, if the probability is greater than $\gamma_c$, the mutation vector replaces the original vector.
\looseness=-1

\noindent\textit{(iv) Selection:}
The differential evolution algorithm employs a greedy approach to determine the superior individual for the next generation, selecting the better one from the current individual and the original parent individual based on the fitness function value. In this paper, the probability is derived directly from the face detector as the fitness score. Hence, a lower probability corresponds to a higher fitness score, signifying a stronger attack, and the individual should be: 
% \begin{equation}
% \label{selection}
% \small
% \theta_{r_1}=\left\{
%          \begin{array}{lr}
%          \theta^p_{r_1} & S^p \leq S^c,\\
         
%          \theta_{r_1} & otherwise,
%          \end{array}
% \right.
% \end{equation}
\begin{equation}
\label{selection}
\small
\theta_{r_1}^{\text{new}}=\left\{
\begin{array}{ll}
\theta_{r_1}^{p}, & S^{p}\le S^{c},\\
\theta_{r_1}^{\text{old}}, & \text{otherwise}.
\end{array}
\right.
\end{equation}
% where $S^p$ represents the probability of the original parent vector and $S^c$ indicates the probability of the current new vector.
where $S^p$ and $S^c$ represent the probabilities of the original parent vector and the current new vector, respectively.

\begin{figure}[t]
        \centering
        \includegraphics[width=\linewidth]{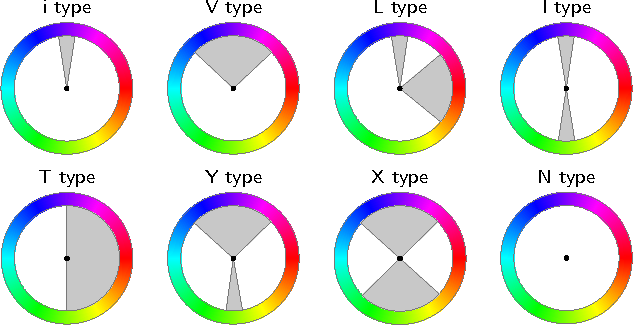}
        \vspace{-0.2in}
        \caption{Harmonic templates.}
        \vspace{-0.2in}
        \label{color harmonization}
\end{figure}

\paragraph{Color Harmonization and Gradient Color Generation.} To achieve a natural and inconspicuous color combination, we introduce an additional constraint known as \textit{color harmonization}. As depicted in Fig.~\ref{color harmonization}, each harmonic template exemplifies a harmonious blend of colors. If the colors of an image fall within the gray region of a specific template, they are deemed harmonious. Consequently, we employ a harmonic template as a constraint to ensure that the color combinations are aesthetically pleasing, providing a sense of comfort and reducing suspicion among individuals. After obtaining the optimized color combination in each iteration,  we specifically utilize harmonic templates of type $T$ to impose constraints on the optimized color combinations.

At each iteration, we implement these colors on the mask to produce the gradient color mask for the color combination. Specifically, we ensure that each row of the mask has the same color, that is, the color gradients from the first row of the mask to another color in the last row are as follows:
\begin{equation}
\label{gradient color}
\small
C_n = C_1 +\frac{(h_n - h_1)}{(h_2 - h_1)}* (C_2 - C_1),
\end{equation}
where the $C_n$ indicates the color in the $n-th$ row, the $C_1$ and $C_2$ represent the two colors in the optimized color combination, $h_1$ is the height of the first row of the mask and $h_2$ is the last row of the mask, $h_n$ is the height of $n-th$ row of the mask. 
The gradient color for each row is computed using the methodology described above. 

In each iteration, we derive the optimal color combination, resulting in an optimal masked face image. To further reduce the gap between the digital and physical domains,  we employ Expectation over Transformation (EOT)~\cite{athalye2018synthesizing}. Specifically, at each iteration, We randomly add some disturbances to the mask, such as adding random noise, brightness adjustment, and randomly perturbing contrast as:
\looseness=-1
\begin{equation}
\label{mask_EOT}
\small
X_{\mathrm{adv}} = \mathbb{E}_{t\sim T}^{1}(\hat{X}_{\mathrm{adv}},\theta)
\end{equation}
where $\hat{X_{adv}}$ represents the perturbed image without EOT. $\theta$ indicates the parameters of color combination, $T$ is the distribution of transformation, and ${X_{adv}}$ is the perturbed image with EOT.
This image is subsequently utilized as input for the visual face detector, yielding the corresponding confidence. Importantly, the color modification does not affect the infrared image; thus, the masked infrared image remains unaltered and serves as the input for the infrared face detector.

% \begin{figure}[t]
%         \centering
%         \includegraphics[width=\linewidth]{pictures/methodology/color harmonization.png}
%         \vspace{-0.2in}
%         \caption{Harmonic templates.}
%         \vspace{-0.2in}
%         \label{color harmonization}
% \end{figure}

\subsection{Band-Aid Optimization}
\label{subsec:band-aid_optimization}
\paragraph{Valid Facial Area.} The subsequent step entails the application of a Band-Aid to the upper portion of the face. Band-Aids, being commonplace facial stickers, possess the advantage of not drawing attention. 
This is primarily attributed to the abundance of facial feature points present in the upper half of the face. By concealing this region, the face detector can be effectively misled. 
Specifically, we also leverage the aforementioned DE algorithm to optimize the parameters $\phi$ of the Band-Aid, including locations and angles. 
% It should be noted that in the real world, it is unreasonable to put band-aids on key parts such as the eyes, as it may even arouse suspicion from others. Therefore, key parts need to be excluded from the search space. Therefore, we generate an effective face-pasting area as the search space, which contains all possible pasting locations for the band-aid stickers.
Since placing band-aids on key regions such as the eyes is unrealistic and may arouse suspicion, we exclude these regions and define an effective face-pasting area as the search space, containing all valid locations for the band-aid stickers.

\paragraph{Location Optimization.} 
The optimal parameters were determined by optimizing the position of the band-aid. However, simply affixing a band-aid directly onto the face is inadequate due to the significant disparities between the virtual and physical domains. While a band-aid might function effectively in the digital domain, its efficacy notably decreases when applied in the physical domain. To mitigate the differences and bridge this gap, it is essential to replicate real physical constraints within the digital environment, including accommodating the curvature of real faces, along with the bending and rotation of band-aid stickers.
\looseness=-1

\begin{figure}[t]
\centering
    \begin{subfigure}[b]{0.155\linewidth}
         \centering
         \includegraphics[width=\linewidth]{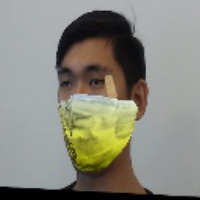}
         %\vspace{-0.25in}
         \caption{}
         %\label{fig:angle}
    \end{subfigure}
    \begin{subfigure}[b]{0.155\linewidth}
         \centering
         \includegraphics[width=\linewidth]{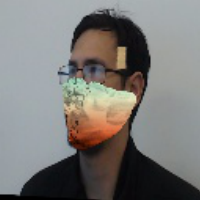}
         %\vspace{-0.25in}
         \caption{}
         %\label{fig:light}
    \end{subfigure}
    \begin{subfigure}[b]{0.155\linewidth}
         \centering
         \includegraphics[width=\linewidth]{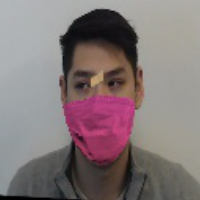}
         %\vspace{-0.25in}
         \caption{}
         %\label{fig:distance}
    \end{subfigure}
    \begin{subfigure}[b]{0.155\linewidth}
         \centering
         \includegraphics[width=\linewidth]{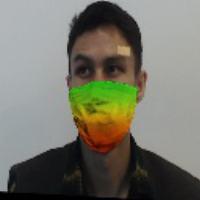}
         %\vspace{-0.25in}
         \caption{}
         %\label{fig:distance}
    \end{subfigure}
    \begin{subfigure}[b]{0.155\linewidth}
         \centering
         \includegraphics[width=\linewidth]{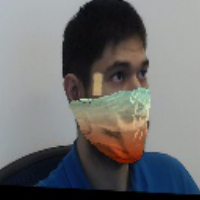}
         %\vspace{-0.25in}
         \caption{}
         %\label{fig:angle}
    \end{subfigure}
    \begin{subfigure}[b]{0.155\linewidth}
         \centering
         \includegraphics[width=\linewidth]{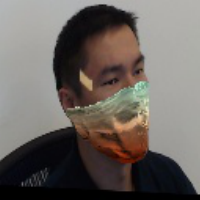}
         %\vspace{-0.25in}
         \caption{}
         %\label{fig:light}
    \end{subfigure}
    \hfill
    \begin{subfigure}[b]{0.155\linewidth}
         \centering
         \includegraphics[width=\linewidth]{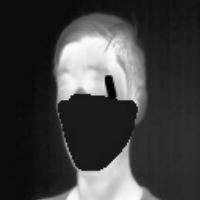}
         %\vspace{-0.25in}
         \caption{}
         %\label{fig:distance}
    \end{subfigure}
    \begin{subfigure}[b]{0.155\linewidth}
         \centering
         \includegraphics[width=\linewidth]{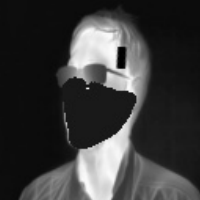}
         %\vspace{-0.25in}
         \caption{}
         %\label{fig:distance}
    \end{subfigure}
    \begin{subfigure}[b]{0.155\linewidth}
         \centering
         \includegraphics[width=\linewidth]{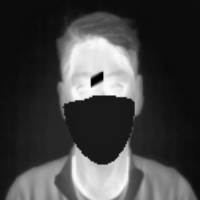}
         %\vspace{-0.25in}
         \caption{}
         %\label{fig:distance}
    \end{subfigure}
    \begin{subfigure}[b]{0.155\linewidth}
         \centering
         \includegraphics[width=\linewidth]{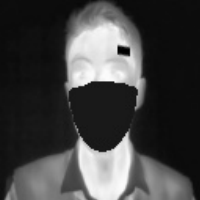}
         %\vspace{-0.25in}
         \caption{}
         %\label{fig:distance}
    \end{subfigure}
    \begin{subfigure}[b]{0.155\linewidth}
         \centering
         \includegraphics[width=\linewidth]{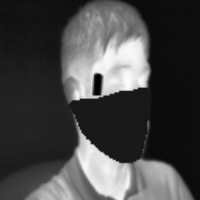}
         %\vspace{-0.25in}
         \caption{}
         %\label{fig:distance}
    \end{subfigure}
    \begin{subfigure}[b]{0.155\linewidth}
         \centering
         \includegraphics[width=\linewidth]{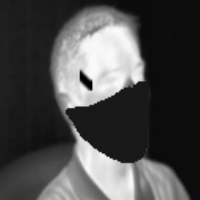}
         %\vspace{-0.25in}
         \caption{}
         %\label{fig:distance}
    \end{subfigure}
    \vspace{-0.1in}
    \caption{Images with visual and infrared patches.}
    \vspace{-0.2in}
    \label{fig:patches_images1}
\end{figure}

\paragraph{3D Transformation.} To accurately simulate these real-world deformations, we start by generating a 3D model of the face from a provided 2D image. Subsequently, to realize the appropriately warped band-aid sticker, we calculate the coordinate transformation matrix, considering both bending and rotation. 
% Using the transformed coordinates, we employ bilinear interpolation~\cite{kirkland2010bilinear} and backward mapping techniques~\cite{se2002vision} to accurately determine the pixel values at each point on the warped band-aid, ensuring a realistic representation in both digital and physical domains.
%
%
During this process, we further enhance our model by integrating the EOT technique, which simulates various real-world scenarios. Specifically, each iteration considers potential inaccuracies in placing the Band-Aid sticker as initially planned. Such errors may result in the Band-Aid not adhering to the intended location. To mitigate these discrepancies, we introduce random perturbations to the optimized parameters, specifically the position and angle of the Band-Aid sticker. This adjustment allows the sticker to shift randomly within a specified vicinity as follows:
\begin{equation}
\label{mask EOT}
\small
X_{adv} = \mathbb{E}_{t\sim T}^{2}(\hat{X_{adv}},\phi),
\end{equation}
where $\hat{X_{adv}}$ represents the perturbed image without EOT. $\phi$ indicates the parameters including location and angle, $T$ is the distribution of transformation, and ${X_{adv}}$ is the perturbed image with EOT.
Given that the masks and band-aids we use are sourced from the physical world, our modifications are limited to altering the color of the mask and the positioning of the band-aids. Consequently, there is no need to account for many of the disparities typically observed between the digital and physical domains, such as TV-loss and NPS-loss. Finally, we successfully generated the optimal adversarial patches in both visual and infrared images, and Fig.~\ref{fig:patches_images1} shows six examples at different orientations and patches.

\section{Evaluation}
\label{sec:evaluation}

\input{tables/table_results_digital_attack_v2}

\begin{table}[t]
\centering
\scriptsize
\setlength{\tabcolsep}{8pt}
\renewcommand{\arraystretch}{0.8}
\caption{ASRs (\%) of infrared images with different methods.}
\vspace{-0.05in}
\label{digital infrared attack}
% \resizebox{\linewidth}{!}{
\begin{tabular}{ccccc}
%{\hsize}{@{}@{\extracolsep{\fill}}cccccc@{}}
\hline
\textbf{Target Model}   & \textbf{AIP $\square$}  & \textbf{HCB $\square$} & \textbf{AdvIB $\blacksquare$} & \textbf{\sysname $\blacksquare$}\\
\hline\hline
\multicolumn{1}{c|}{TFW}  & \cellcolor{mycolor4}92.36 & \cellcolor{mycolor4}96.73 & \cellcolor{mycolor4}92.27   & 
 \cellcolor{mycolor4}\textbf{100.00}   \\
%\hline
\multicolumn{1}{c|}{Yolov8-Face}    & \cellcolor{mycolor1}10.18 & \cellcolor{mycolor3}67.20 & \cellcolor{mycolor3}58.73  & \cellcolor{mycolor4}\textbf{86.18}  \\
%\hline
\multicolumn{1}{c|}{RetinaFace}   & \cellcolor{mycolor4}89.09 & \cellcolor{mycolor4}99.09 & \cellcolor{mycolor4}98.55   & \cellcolor{mycolor4}\textbf{100.00}  \\
%\hline
\multicolumn{1}{c|}{SCRFD}   & \cellcolor{mycolor3}62.73 & \cellcolor{mycolor4}96.36 & \cellcolor{mycolor4}98.73   & \cellcolor{mycolor4}\textbf{99.27}   \\
%\hline
\multicolumn{1}{c|}{MTCNN}    & \cellcolor{mycolor4}78.18& \cellcolor{mycolor4}96.73 & \cellcolor{mycolor4}98.73  & \cellcolor{mycolor4}\textbf{99.09}   \\
%\hline
\multicolumn{1}{c|}{ULFD}    & \cellcolor{mycolor3}65.27 & \cellcolor{mycolor4}90.55& \cellcolor{mycolor4}93.27  & \cellcolor{mycolor4}\textbf{97.27}   \\
%\hline
\multicolumn{1}{c|}{OpenCV}   & \cellcolor{mycolor4}79.27& \cellcolor{mycolor4}99.64& \cellcolor{mycolor4}99.45  & \cellcolor{mycolor4}\textbf{100.00}  \\
\hline
\end{tabular}
\vspace{-0.15in}
% }
% \caption{ASRs (\%) results of the attack in the infrared images with various models and methods.}
% \label{digital infrared attack}
\end{table}

% \vspace{-0.2in}
\begin{table}[t]
\centering
\scriptsize
\setlength{\tabcolsep}{10pt}
\renewcommand{\arraystretch}{0.8}
\caption{Physical attack ASRs (\%).}
\vspace{-0.05in}
\label{Physical infrared attack}
% \resizebox{\linewidth}{!}{
\begin{tabular}{ccccc}
\hline

\textbf{Target Model}& \textbf{No Mask}  & \textbf{AdvIB} &\textbf{HCB}& \textbf{\sysname} \\ \hline\hline
\multicolumn{1}{c|}{TFW} 
& \cellcolor{mycolor1}0.00
& \cellcolor{mycolor4}100.00   
& \cellcolor{mycolor4}100.00
& \cellcolor{mycolor4}100.00 
\\ %\hline
\multicolumn{1}{c|}{Yolov8-Face}  
&\cellcolor{mycolor1}0.00
&\cellcolor{mycolor1}21.11
&\cellcolor{mycolor1}5.00
&\cellcolor{mycolor4}97.27      \\ %\hline
\multicolumn{1}{c|}{MTCNN}        
&\cellcolor{mycolor1}0.00
&\cellcolor{mycolor3}71.11 
&\cellcolor{mycolor2}47.50
&\cellcolor{mycolor4}97.27      \\ %\hline
\multicolumn{1}{c|}{OpenCV}  
&\cellcolor{mycolor1}0.00
&\cellcolor{mycolor4}80.00 
&\cellcolor{mycolor3}70.00
&\cellcolor{mycolor4}100.00
\\ \hline
\end{tabular}
\vspace{-0.2in}
% }
% \caption{Physical attack ASRs (\%).}
% \label{Physical infrared attack}
\end{table}
\begin{table}[t]
\centering
\scriptsize
\setlength{\tabcolsep}{20pt}
\renewcommand{\arraystretch}{0.8}
\caption{ASRs (\%) solely on visual detectors.}
\label{tab:only_consider_visual}
\vspace{-0.05in}
% \resizebox{\linewidth}{!}{
\begin{tabular}{lcc}
%{\hsize}{@{}@{\extracolsep{\fill}}cccccccccccccc@{}}
\hline
\textbf{Target Model} & \textbf{Only visual}  & \textbf{Fused}   \\
\hline\hline
\multicolumn{1}{c|}{Yolov5-Face} & \cellcolor{mycolor4}76.47  & \cellcolor{mycolor4}\textbf{92.72}     \\
%\hline
\multicolumn{1}{c|}{Yolov8-Face} & \cellcolor{mycolor3}68.56 & \cellcolor{mycolor4}\textbf{100.00}   \\
%\hline 
\multicolumn{1}{c|}{RetinaFace} & \cellcolor{mycolor3}72.73 & \cellcolor{mycolor3}\textbf{99.09}    \\
%\hline
\multicolumn{1}{c|}{SCRFD} & \cellcolor{mycolor3}66.59 & \cellcolor{mycolor4}\textbf{100.00}   \\
%\hline
\multicolumn{1}{c|}{MTCNN} & \cellcolor{mycolor3}71.48 & \cellcolor{mycolor4}\textbf{92.91}   \\
%\hline
\multicolumn{1}{c|}{ULFD} & \cellcolor{mycolor4}78.67& \cellcolor{mycolor4}\textbf{82.55}   \\
%\hline
\multicolumn{1}{c|}{MogFace} & \cellcolor{mycolor4}77.56& \cellcolor{mycolor4}\textbf{85.71}  \\
%\hline
\multicolumn{1}{c|}{Dlib} & \cellcolor{mycolor3}71.36& \cellcolor{mycolor4}\textbf{100.00}  \\
%\hline
\multicolumn{1}{c|}{OpenCV} & \cellcolor{mycolor3}65.72& \cellcolor{mycolor4}\textbf{98.36}  \\
\hline
\end{tabular}
\vspace{-0.15in}
% }
% \caption{ASRs (\%) by optimizing adversarial patches solely on visual face detectors, compared to optimizations on both visual and infrared face detectors.}
% \label{tab:only_consider_visual}
\end{table}

\begin{table}[t]
\centering
\scriptsize
% \resizebox{\linewidth}{!}{
\setlength{\tabcolsep}{20pt}
\renewcommand{\arraystretch}{0.8}
\caption{ASRs (\%) solely on infrared detectors.}
\label{only consider infrared}
\vspace{-0.05in}
% \resizebox{\linewidth}{!}{
\begin{tabular}{lcc}
%{\hsize}{@{}@{\extracolsep{\fill}}cccccc@{}}
\hline
\textbf{Target Model}   & \textbf{Only Infrared}  & \textbf{Fused} \\
\hline\hline
\multicolumn{1}{c|}{TFW}  & \cellcolor{mycolor3}60.92 & \cellcolor{mycolor4}\textbf{100.00}     \\
%\hline
\multicolumn{1}{c|}{Yolov8-Face}    & \cellcolor{mycolor3}71.09 & \cellcolor{mycolor4}\textbf{86.18}   \\
%\hline
\multicolumn{1}{c|}{RetinaFace}   & \cellcolor{mycolor3}62.01 & \cellcolor{mycolor4}\textbf{100.00}    \\
%\hline
\multicolumn{1}{c|}{SCRFD}   & \cellcolor{mycolor3}61.57 & \cellcolor{mycolor4}\textbf{99.27}       \\
%\hline
\multicolumn{1}{c|}{MTCNN}    & \cellcolor{mycolor3}69.35& \cellcolor{mycolor4}\textbf{99.09}      \\
%\hline
\multicolumn{1}{c|}{ULFD}    & \cellcolor{mycolor3}63.54 & \cellcolor{mycolor4}\textbf{97.27}    \\
%\hline
\multicolumn{1}{c|}{OpenCV}   & \cellcolor{mycolor3}61.46& \cellcolor{mycolor4}\textbf{100.00}   \\
\hline
\end{tabular}
\vspace{-0.2in}
% }
% \caption{ASRs (\%) by optimizing adversarial patches solely on infrared face detectors, compared to optimizations on both visual and infrared face detectors.}
% \label{only consider infrared}
\end{table}

\subsection{Experiment Setup}
\label{subsec:experiment_setup}

\paragraph{Datasets.}
We used an open-source dataset collected and released by SPEAKINGFACES ~\cite{abdrakhmanova2021speakingfaces} containing both visual and infrared images captured by thermal cameras.
Specifically, this open-source dataset consists of 2,556 pairs of thermal-visual images collected from 142 subjects with manually annotated face-bounding boxes.
\looseness=-1

\paragraph{Evaluated Models.}
We select nine commonly used and SOTA visual face detection models: Yolov5-Face~\cite{qi2022yolo5face}, Yolov8-Face~\cite{der2023yolov8} RetinaFace~\cite{deng2019retinaface}, SCRFD~\cite{guo2021sample}, MTCNN~\cite{zhang2016joint}, MogFace~\cite{liu2022mogface}, ULFD~\cite{lin2019ulfd}, Dlib~\cite{king2009dlib}, OpenCV~\cite{bradski2000opencv}. 
All these models can detect human faces with masks.
For the infrared face,  we utilize the SOTA infrared face detectors: TFW~\cite{kuzdeuov2022tfw}, Yolov8-Face~\cite{qi2022yolo5face}, RetinaFace~\cite{deng2019retinaface}, SCRFD~\cite{guo2021sample}, MTCNN~\cite{zhang2016joint}, ULFD~\cite{lin2019ulfd}, and OpenCV~\cite{bradski2000opencv}, which have the ability to successfully detect and accurately recognize faces in infrared images. 
% To evaluate the effectiveness of the infrared patch, we employed face detection algorithms to assess its efficacy.

\paragraph{Evaluation Metrics.} For both white-box and black-box attacks, we adopt the Attack Success Rate (ASR) as the evaluation metric, which denotes the ratio of successful spoofing trials among all testing trials.
 Our objective is to transform true positive (TP) cases into false negative (FN) cases when scanned by visual-infrared face detectors.

\paragraph{Experiment Details.}
For the visual image, we conducted experiments comparing different SOTA methods. In practice, we consider three white-box attacks, Adv-Patch~\cite{Adv-Patch}, Adv-Cloak~\cite{Adv-Cloak}, Adv-Texture~\cite{Adv-Texture}, and two black-box attacks, HardBeat~\cite{hardbeat} and RS-Patch~\cite{rs-patch}. 
To ensure the fairness of the comparative experiments, we evaluate white-box attacks in black-box scenarios. 
Consequently, we conduct black-box attacks to assess the performance of the white-box methods in the black-box scenarios. 
In particular, the white-box attack methods~\cite{Adv-Patch, Adv-Cloak, Adv-Texture} are trained using the Yolov5-Face model and tested on other models. 
Therefore, we utilize finite difference methods to estimate the gradient of each pixel. We set the max query $Q=10000$, $\gamma_m = 0.5$, $\gamma_c = 0.6$, the harmonic templates that we choose is the $T$ type. For the facial mask generation, we generate the facial mask by following FMA-3D \cite{wang2021facex}. For 3D transformation, we implement 3DMM~\cite{blanz2023morphable} to generate the 3D face model. To generate the warped band-aid sticker, the implemented method follows adv-sticker~\cite{adv-sticker}.

For the infrared image, we also conducted comparison experiments with the SOTA methods, including two white-box attacks, AIP~\cite{AIP} and HCB~\cite{HCB}, and one black-box attack AdvIB~\cite{AdvIB}. AIP is also trained using the TFW model and evaluated on other models. The number of hot and cold blocks used in HCB~\cite{HCB} and AdvIB~\cite{AdvIB} is $4$. Note that the previous work only needs to attack either the visual model or the infrared model, while \sysname attacks both the visual and infrared models simultaneously.

\looseness=-1

\paragraph{Experiment Devices.} Our method is implemented in PyTorch 1.11.0+cu113 on one NVIDIA RTX A6000 GPU. 
In our physical experimental setup, 
we utilize the Guide Sensmart MobIR 2T as the infrared camera, which is a VOx infrared detector with a resolution of $256 \times 192$ pixels and a \SI{25}{\hertz} frame rate. It has a \SI{3.2}{\milli\meter} lens with a $56^\circ$ field of view and can measure temperatures ranging from 
\SI{-20}{\degreeCelsius} to \SI{150}{\degreeCelsius} 
with an accuracy of $\pm$\SI{2}{\degreeCelsius}
or $2\%$. The device weighs less than \SI{35}{\gram} and operates with a power supply of $4.5$--\SI{5.5}{\volt}.
For the visual camera, we utilized the Apple iPhone 13 Pro Max smartphone.
We recorded videos at 30 frames per second (FPS) under varying light intensities, distances, and angles. Each video recording lasted 20 seconds, resulting in 600 frames per video.
In practice, we obtained the optimized parameters, including the mask color, the band-aid angle, and location, then printed the gradient color on paper, as well as attached the mask and band-aid to the face.

\subsection{Digital Domain Attack} 
\label{subsec:digital_domain_attack}

\paragraph{Different Visual Face Detection Models.} To evaluate the performance of \sysname against various  visual face detectors, we utilized the TFW infrared face detector~\cite{kuzdeuov2022tfw} and various visual face detectors. Table~\ref{tab:digital attack visual domain} shows the ASRs for both white-box and black-box methods.
For the white-box methods, the average ASRs for Adv-Patch, Adv-Cloak, and Adv-Texture are $64.25\%$, $50.78\%$, and $52.94\%$, respectively. Notably, transfer attacks on YOLOv8-Face exhibit high ASRs, reaching $92.17\%$, $89.82\%$, $87.82\%$, and $95.80\%$, respectively. In contrast, the ASRs for MTCNN are remarkably low, with values of $10.18\%$, $8.91\%$, $10.36\%$, and $13.09\%$, respectively. These results underscore the limited transferability in black-box attack scenarios.

For the black-box methods, the average ASRs of HardBeat, RS-Patch, and \sysname are $86.95\%$, $93.13\%$ and $94.59\%$, respectively. It is worth noting that Hardbeat and RS-Patch slightly outperform~\sysname on some models, such as OpenCV. This discrepancy arises because these methods focus exclusively on performance in the digital domain, which significantly differs from the physical domain. In contrast, we consider various physical constraints to reduce the gap between the digital and physical domains, narrowing the range of variables such as patch color and sticker position. 
Moreover, the ASR of ULFD and Mogface are both below $90\%$, which is lower than those of other models. This discrepancy arises because these two models are more complex, making them more challenging to optimize and consequently harder to find the optimal value. Nevertheless, \sysname demonstrates the highest ASR achieved among all methods, even the lowest ASR of $82.55\%$ remains at a high level. 
\looseness=-1

\begin{figure*}[t]
\centering
    \begin{subfigure}[b]{0.325\linewidth}
         \centering
         \includegraphics[width=\linewidth]{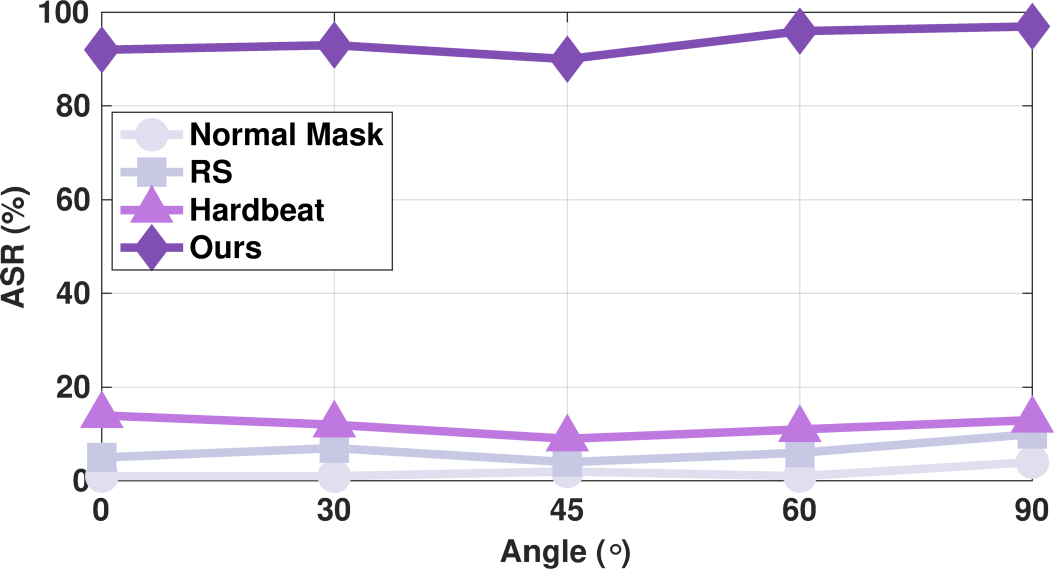}
         % \vspace{-0.2in}
         \caption{Different angles.}
         \label{fig:angle}
    \end{subfigure}
    \begin{subfigure}[b]{0.325\linewidth}
         \centering
         \includegraphics[width=\linewidth]{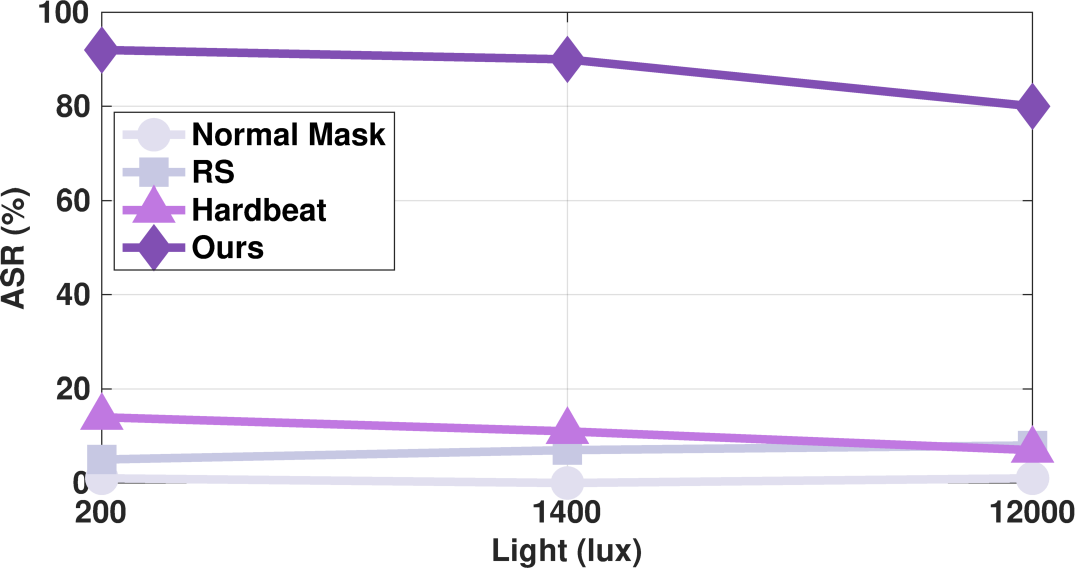}
         % \vspace{-0.2in}
         \caption{Different light conditions.}
         \label{fig:light}
    \end{subfigure}
    \begin{subfigure}[b]{0.325\linewidth}
         \centering
         \includegraphics[width=\linewidth]{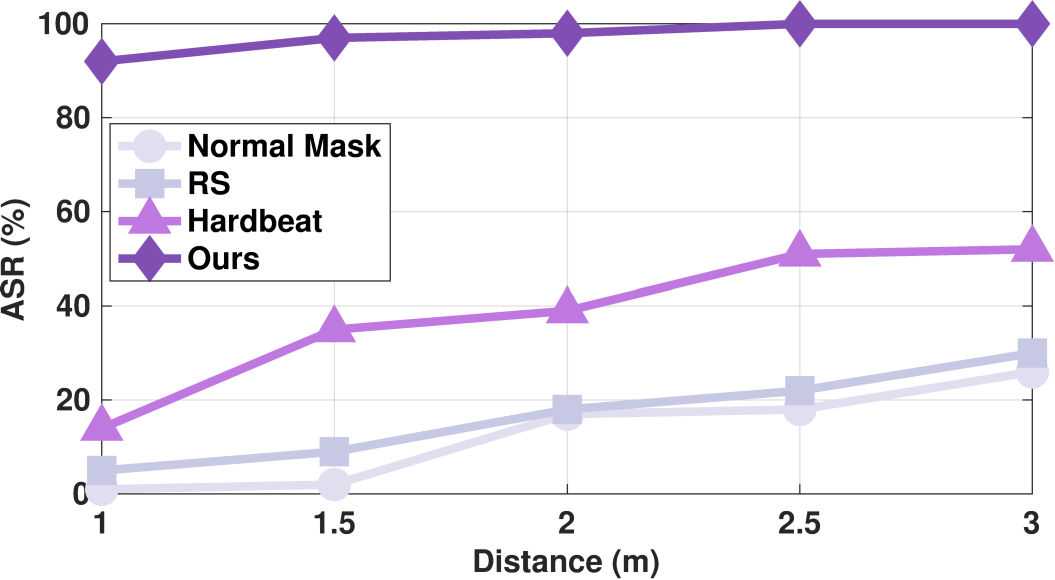}
         % \vspace{-0.2in}
         \caption{Different distances.}
         \label{fig:distance}
    \end{subfigure}
    \vspace{-0.05in}
    \caption{\sysname's ASRs at different angles, light conditions and distances.}
    \vspace{-0.2in}
    \label{fig:physical_visual_experiments}
\end{figure*}

\begin{table}[t]
\centering
\scriptsize
%\resizebox{\linewidth}{!}{
\setlength{\tabcolsep}{12pt}
\renewcommand{\arraystretch}{0.8}
\caption{ASRs (\%) with only mask or band-aid.}
\vspace{-0.05in}
% \resizebox{\linewidth}{!}{
\begin{tabular}{lccc}
%{\hsize}{@{}@{\extracolsep{\fill}}cccccccccccccc@{}}
\hline

\textbf{Target Model} & \textbf{Only Mask}  & \textbf{Only Band-Aid}& \textbf{Both}   \\
\hline\hline
\multicolumn{1}{c|}{Yolov5-Face} & \cellcolor{mycolor4}89.40  & \cellcolor{mycolor1}23.09 & \cellcolor{mycolor4}\textbf{92.72}     \\
%\hline
\multicolumn{1}{c|}{Yolov8-Face} & \cellcolor{mycolor4}86.55 & \cellcolor{mycolor4}90.91& \cellcolor{mycolor4}\textbf{100.00}    \\
%\hline 
\multicolumn{1}{c|}{RetinaFace} & \cellcolor{mycolor4}85.95 & \cellcolor{mycolor3}62.55& \cellcolor{mycolor4}\textbf{99.09}     \\
%\hline
\multicolumn{1}{c|}{SCRFD} & \cellcolor{mycolor4}83.66 & \cellcolor{mycolor2}26.36& \cellcolor{mycolor4}\textbf{100.00}    \\
%\hline
\multicolumn{1}{c|}{MTCNN} & \cellcolor{mycolor4}84.37 & \cellcolor{mycolor1}1.27& \cellcolor{mycolor4}\textbf{92.91}    \\
%\hline
\multicolumn{1}{c|}{ULFD} & \cellcolor{mycolor4}76.84& \cellcolor{mycolor1}16.18& \cellcolor{mycolor4}\textbf{82.55}    \\
%\hline
\multicolumn{1}{c|}{MogFace} & \cellcolor{mycolor3}71.35& \cellcolor{mycolor3}36.91& \cellcolor{mycolor4}\textbf{85.71}   \\
%\hline
\multicolumn{1}{c|}{Dlib} & \cellcolor{mycolor4}88.26& \cellcolor{mycolor2}17.09& \cellcolor{mycolor4}\textbf{100.00}   \\
%\hline
\multicolumn{1}{c|}{OpenCV} & \cellcolor{mycolor4}84.67& \cellcolor{mycolor2}\textbf{42.18} & \cellcolor{mycolor4}\textbf{98.36}  \\
\hline
\end{tabular}
\vspace{-0.15in}
% }
% \caption{ASRs (\%) obtained from employing only the mask, only the band-aid sticker, and an optimization strategy that incorporates both the mask and the band-aid sticker.}
\label{only mask or band-aid}
\end{table}

\paragraph{Different Infrared Face Detection Models.}
To verify the robustness of our method against diverse infrared face detectors, we maintained the visual face detector as Yolov8-Face~\cite{der2023yolov8} and sequentially substituted the infrared face detectors with different models for evaluation. Similarly, the white-box method AIP was trained on TFW and evaluated on other models to test its ability to attack the black-box models. 
Table~\ref{digital infrared attack} lists the ASRs of different methods when targeting different infrared face detectors. For the white-box methods AIP and HCB, the highest ASRs are $92.36\%$ and $99.64\%$, while the lowest ASRs are $10.18\%$ and $67.20\%$, respectively; their average ASRs are $68.15\%$ and $92.33\%$, respectively, indicating the first method has limited transferability in black-box scenarios. In contrast, for the black-box attacks AdvIB and our proposed method, the average ASRs are $91.39\%$ and $97.40\%$, respectively. Notably, even the lowest ASR of our method when targeting Yolov8-Face, $86.18\%$, remains at a high level. This highlights the effectiveness of the patches generated by our method, as they outperform comparable methods in achieving the highest ASR
\looseness=-1

% % \vspace{-0.2in}
% \begin{table}[t]
% \centering
% \scriptsize
% \setlength{\tabcolsep}{10pt}
% \renewcommand{\arraystretch}{0.8}
% \caption{Physical attack ASRs (\%).}
% \vspace{-0.05in}
% \label{Physical infrared attack}
% % \resizebox{\linewidth}{!}{
% \begin{tabular}{ccccc}
% \hline

% \textbf{Target Model}& \textbf{No Mask}  & \textbf{AdvIB} &\textbf{HCB}& \textbf{\sysname} \\ \hline\hline
% \multicolumn{1}{c|}{TFW} 
% & \cellcolor{mycolor1}0.00
% & \cellcolor{mycolor4}100.00   
% & \cellcolor{mycolor4}100.00
% & \cellcolor{mycolor4}100.00 
% \\ %\hline
% \multicolumn{1}{c|}{Yolov8-Face}  
% &\cellcolor{mycolor1}0.00
% &\cellcolor{mycolor1}21.11
% &\cellcolor{mycolor1}5.00
% &\cellcolor{mycolor4}97.27      \\ %\hline
% \multicolumn{1}{c|}{MTCNN}        
% &\cellcolor{mycolor1}0.00
% &\cellcolor{mycolor3}71.11 
% &\cellcolor{mycolor2}47.50
% &\cellcolor{mycolor4}97.27      \\ %\hline
% \multicolumn{1}{c|}{OpenCV}  
% &\cellcolor{mycolor1}0.00
% &\cellcolor{mycolor4}80.00 
% &\cellcolor{mycolor3}70.00
% &\cellcolor{mycolor4}100.00
% \\ \hline
% \end{tabular}
% \vspace{-0.15in}
% % }
% % \caption{Physical attack ASRs (\%).}
% % \label{Physical infrared attack}
% \end{table}

\subsection{Physical Domain Attack} 
\label{subsec:physical_domain_attack}

\paragraph{Visual Image.}
In a physical domain attack, we first aim to obtain the optimized parameters, including the mask color, location, and angle of the band-aid. To verify the ASR of the patch in different scenarios in the real physical world, we conducted three sets of experiments following the previous work \cite{Adv-Patch, Adv-Cloak, Adv-Texture, adv-mask, adv-sticker} on different angles, distances, and light intensities, shown in Fig.~\ref{fig:physical_visual_experiments}. For comparative purposes, we evaluate the ASR under conditions where participants (IRB approved) wear standard surgical masks. Subsequently, we print out the patches generated using the HardBeat algorithm~\cite{hardbeat} and RS-Patch ~\cite{rs-patch}, both have achieved the highest ASRs within the digital domain.
\looseness=-1

Fig.~\ref{fig:angle} demonstrates the ASRs in the situations of different angles. For HardBeat, ASRs at angles of $0^\circ$, $30^\circ$, $45^\circ$, $60^\circ$, and $90^\circ$ are recorded at $14.29\%$, $12.45\%$, $9.37\%$, $11.91\%$, and $13.75\%$, respectively. For RS-Patch, the corresponding ASRs are $9.38\%$, $5.82\%$, $4.27\%$, $6.09\%$, and $10.25\%$. In contrast, our method exhibits ASRs of $92.72\%$, $93.55\%$, $90.71\%$, $96.36\%$, and $97.18\%$, respectively. The performance of HardBeat and RS-Patch methods is contingent upon a predetermined assumption regarding the camera's position to ensure the camera captures the adversarial patch completely. Consequently, as the angle increases, their effectiveness diminishes. Conversely, the ASR of our method remains consistently high across varying angles, demonstrating that our method does not require prior knowledge of the camera's position, thereby proving to be more robust and applicable in real-world scenarios.
\looseness=-1

Fig.~\ref{fig:light} offers a comparative analysis of ASRs under three lighting conditions: An illuminance of $200\,lux$ corresponds to indoor settings, $1400\, lux$ to cloudy outdoor conditions, and $12,000\, lux$ to sunny outdoor environments. Specifically, for the HardBeat method, ASRs at light intensities of $200\,lux$, $1400\,lux$, and $12000\,lux$ are $14.16\%$, $11.95\%$, and $7.36\%$, respectively. For the RS-Patch method, the ASRs are recorded at $5.78\%$, $7.45\%$, and $8.92\%$. In comparison, our method demonstrates ASRs of $92.64\%$, $90.55\%$, and $80.29\%$. These results underscore the superior performance of our algorithm, which can be ascribed to its efficient bridging of the gap between the digital and physical domains. 
% A noteworthy trend is the inverse correlation between light intensity and ASR, suggesting that increased brightness may impede the effectiveness of the attack due to the adverse effects of glare and overexposure on the adversarial patch's detectability.
\looseness=-1

% \begin{figure*}[t]
% \centering
%     \begin{subfigure}[b]{0.325\linewidth}
%          \centering
%          \includegraphics[width=\linewidth]{pictures/experiment/asr_angle.pdf}
%          % \vspace{-0.2in}
%          \caption{Different angles.}
%          \label{fig:angle}
%     \end{subfigure}
%     \begin{subfigure}[b]{0.325\linewidth}
%          \centering
%          \includegraphics[width=\linewidth]{pictures/experiment/asr_light.pdf}
%          % \vspace{-0.2in}
%          \caption{Different light conditions.}
%          \label{fig:light}
%     \end{subfigure}
%     \begin{subfigure}[b]{0.325\linewidth}
%          \centering
%          \includegraphics[width=\linewidth]{pictures/experiment/asr_distance.pdf}
%          % \vspace{-0.2in}
%          \caption{Different distances.}
%          \label{fig:distance}
%     \end{subfigure}
%     \vspace{-0.1in}
%     \caption{\sysname's ASRs at different angles, light conditions and distances.}
%     \vspace{-0.2in}
%     \label{fig:physical_visual_experiments}
% \end{figure*}

Fig.~\ref{fig:distance} illustrates the relationship between the effectiveness of attacks and the distance from the patch to the camera, ranging from 1 to 3 meters. At distances of $1m$, $1.5m$, $2m$, $2.5m$, and $3m$, the ASRs for the HardBeat method are $14.28\%$, $35.68\%$, $39.86\%$, $51.08\%$, and $52.73\%$, respectively. For the RS-Patch method, the ASRs are $5.18\%$, $9.73\%$, $18.22\%$, $22.16\%$, and $30.24\%$. By contrast, our method exhibits exceptionally high ASRs of $92.51\%$, $96.37\%$, $97.89\%$, $100.00\%$, $100.00\%$, respectively. These results demonstrate that our method far outperforms other methods. Notably, the ASRs for all methods increase with distance because of a trend attributed to the reduced size of the detected face at greater distances, which complicates face detection.
\looseness=-1

\paragraph{Infrared Image.} Similar to visual images, \sysname launching attacks on infrared images also involves initially uploading the infrared images to the digital domain for the attack and subsequently executing the attacks in the physical world. We manually applied a small amount of cold gel beneath the Band-Aid to serve as insulation. For comparative analysis, we selected HCB and AdvIB as benchmark methods due to their robust performance in the digital domain. 
%
% % \vspace{-0.2in}
% \begin{table}[t]
% \centering
% \scriptsize
% \setlength{\tabcolsep}{8pt}
% \renewcommand{\arraystretch}{1.1}
% \caption{Physical attack ASRs (\%).}
% % \vspace{-0.15in}
% \label{Physical infrared attack}
% % \resizebox{\linewidth}{!}{
% \begin{tabular}{ccccc}
% \hline
%
% \textbf{Target Model}& \textbf{No Mask}  & \textbf{AdvIB} &\textbf{HCB}& \textbf{\sysname} \\ \hline\hline
% \multicolumn{1}{c|}{TFW} 
% & \cellcolor{mycolor1}0.00
% & \cellcolor{mycolor4}100.00   
% & \cellcolor{mycolor4}100.00
% & \cellcolor{mycolor4}100.00 
% \\ %\hline
% \multicolumn{1}{c|}{Yolov8-Face}  
% &\cellcolor{mycolor1}0.00
% &\cellcolor{mycolor1}21.11
% &\cellcolor{mycolor1}5.00
% &\cellcolor{mycolor4}97.27      \\ %\hline
% \multicolumn{1}{c|}{MTCNN}        
% &\cellcolor{mycolor1}0.00
% &\cellcolor{mycolor3}71.11 
% &\cellcolor{mycolor2}47.50
% &\cellcolor{mycolor4}97.27      \\ %\hline
% \multicolumn{1}{c|}{OpenCV}  
% &\cellcolor{mycolor1}0.00
% &\cellcolor{mycolor4}80.00 
% &\cellcolor{mycolor3}70.00
% &\cellcolor{mycolor4}100.00
% \\ \hline
% \end{tabular}
% % }
% % \caption{Physical attack ASRs (\%).}
% % \label{Physical infrared attack}
% \end{table}
%
Table~\ref{Physical infrared attack} displays the ASRs for physical attacks on infrared images. The ASRs for AdvIB across the TFW, Yolov8-Face, MTCNN, and OpenCV models are $100.00\%$, $21.11\%$, $71.11\%$, and $80.00\%$, respectively. For HCB, the ASRs are $100.00\%$, $5.00\%$, $47.50\%$, and $70.00\%$. In comparison, our method consistently achieves competitive ASRs of $100.00\%$, $97.27\%$, $97.27\%$, and $100.00\%$. These results demonstrate that the patches generated by our method significantly outperform those from other methods, effectively exploiting vulnerabilities in infrared face detection systems.

\section{Ablation Study}
\label{sec:ablation_study}

\subsubsection{Single Visual or Infrared Detector.}
% Table~\ref{tab:only_consider_visual} presents the ASRs of \sysname in the optimization of a patch based solely on visual face detectors. The ASRs for various detectors: Yolov5-Face, Yolov8-Face, RetinaFace, SCRFD, MTCNN, ULFD, Mogface, Dlib, and OpenCV are $76.47\%$, $68.56\%$, $72.73\%$, $66.59\%$, 
% $71.48\%$, $78.67\%$, $77.56\%$, and $65.72\%$, respectively. In contrast, patches optimized based on visual-infrared fused face detectors exhibit significantly higher ASRs, namely $92.72\%$, $100.00\%$, $99.09\%$, $100.00\%$, $92.91\%$, $82.55\%$, $85.71\%$, $100.00\%$, and $98.36\%$. 
% Table~\ref{only consider infrared} shows ASRs for patches optimized solely on infrared detectors, including TFW, Yolov8-Face, RetinaFace, SCRFD, MTCNN, ULFD, and OpenCV, are $60.92\%$, $71.09\%$, $62.01\%$, $61.57\%$, $69.35\%$, $63.54\%$, and $61.46\%$, respectively. These results are much lower than those obtained from patches optimized using both visual and infrared information, which yields ASRs of $100.00\%$, $86.18\%$, $100.00\%$, $99.27\%$, $99.09\%$, $97.27\%$, and $100.00\%$. 

Tables~\ref{tab:only_consider_visual} and~\ref{only consider infrared} report the ASRs of \sysname when the patch is optimized on a single modality. Optimizing solely on visual detectors yields ASRs of only 65.72\%--78.67\%, while optimizing solely on infrared detectors gives an even lower range of 60.92\%--71.09\%. In contrast, jointly optimizing over the visual-infrared fused detectors raises the ASR above 82\% on every model, reaching 100\% on several. These results confirm that exploiting both modalities is essential to the effectiveness of \sysname.

\looseness=-1

\paragraph{Single Facial Mask or Band-Aid Sticker.}
% Table~\ref{only mask or band-aid} outlines the ASRs of \sysname when a mask or a band-aid sticker is used or both are used. With only a mask, the ASRs for Yolov5-Face, Yolov8-Face, RetinaFace, SCRFD, MTCNN, ULFD, MogFace, Dlib, and OpenCV are $89.40\%$, $86.55\%$, $85.95\%$, $83.66\%$, $84.37\%$, $76.84\%$, $71.35\%$, $88.26\%$, and $84.67\%$, respectively. In the scenario where only a Band-Aid sticker is employed, the ASRs are $23.09\%$, $90.91\%$, $62.55\%$, $26.36\%$, $1.27\%$, $16.18\%$, $36.91\%$, $17.09\%$, and $42.18\%$. 

Table~\ref{only mask or band-aid} reports the ASRs of \sysname when only the mask, only the band-aid, or both are used. Using only the mask already achieves consistently high ASRs (71.35\%--89.40\%), whereas using only the band-aid is far less stable, ranging from 1.27\% to 90.91\% across detectors. Combining both consistently yields the highest ASRs, showing that the two components complement each other and are jointly essential to the attack.

\section{Conclusion}
% This paper presents \sysname, an inconspicuous adversarial patch designed for visual-infrared fused face detectors, which are widely utilized in temperature screening systems and face-tracking surveillance systems.
% By meticulously optimizing the parameters of the mask and band-aid sticker, we effectively launch attacks on the target model in the physical domain. Extensive experimentation demonstrates the profoundly damaging nature of the generated patch, posing a significant threat to applications such as temperature screening systems. 
% Furthermore, our defense experiments demonstrated that enhancing robustness is a highly effective measure.
This paper presents \sysname, an inconspicuous adversarial patch against visual-infrared fused face detectors, which are widely deployed in temperature screening and face-tracking surveillance systems. By jointly optimizing a gradient-color mask and a band-aid sticker, \sysname reliably bypasses the target models in the physical world. Extensive experiments demonstrate its effectiveness across diverse conditions, revealing a significant vulnerability in real-world visual-infrared face detection systems.
\looseness=-1

% \section*{Acknowledgements}
% Please insert your acknowledgments here.

\bibliography{main}
\end{document}

%% file: tables/table_results_digital_attack_v2.tex
\begin{table*}[t]
\centering
\scriptsize
\setlength{\tabcolsep}{6pt}
\renewcommand{\arraystretch}{0.8}
\caption{ASRs (\%) of \sysname's digital attack in visual images. Specifically, the white-box ($\square$) attack methods are trained on the Yolov5-Face model and tested on other models, and the black-box ($\blacksquare$) attack methods are directly tested on all models. ASR color bricks in tables: \includegraphics[width=8pt]{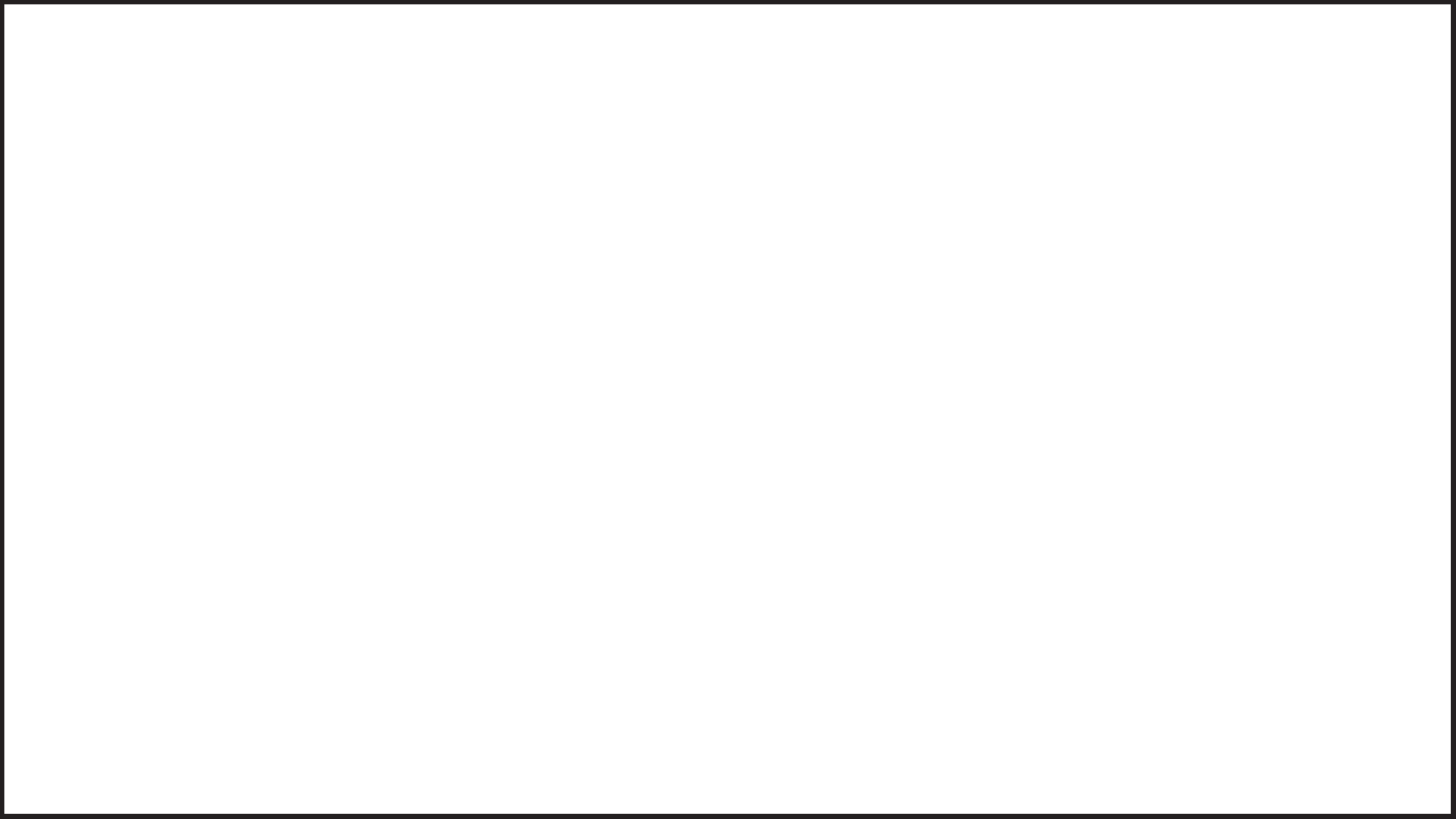} : 0--25\%, \includegraphics[width=8pt]{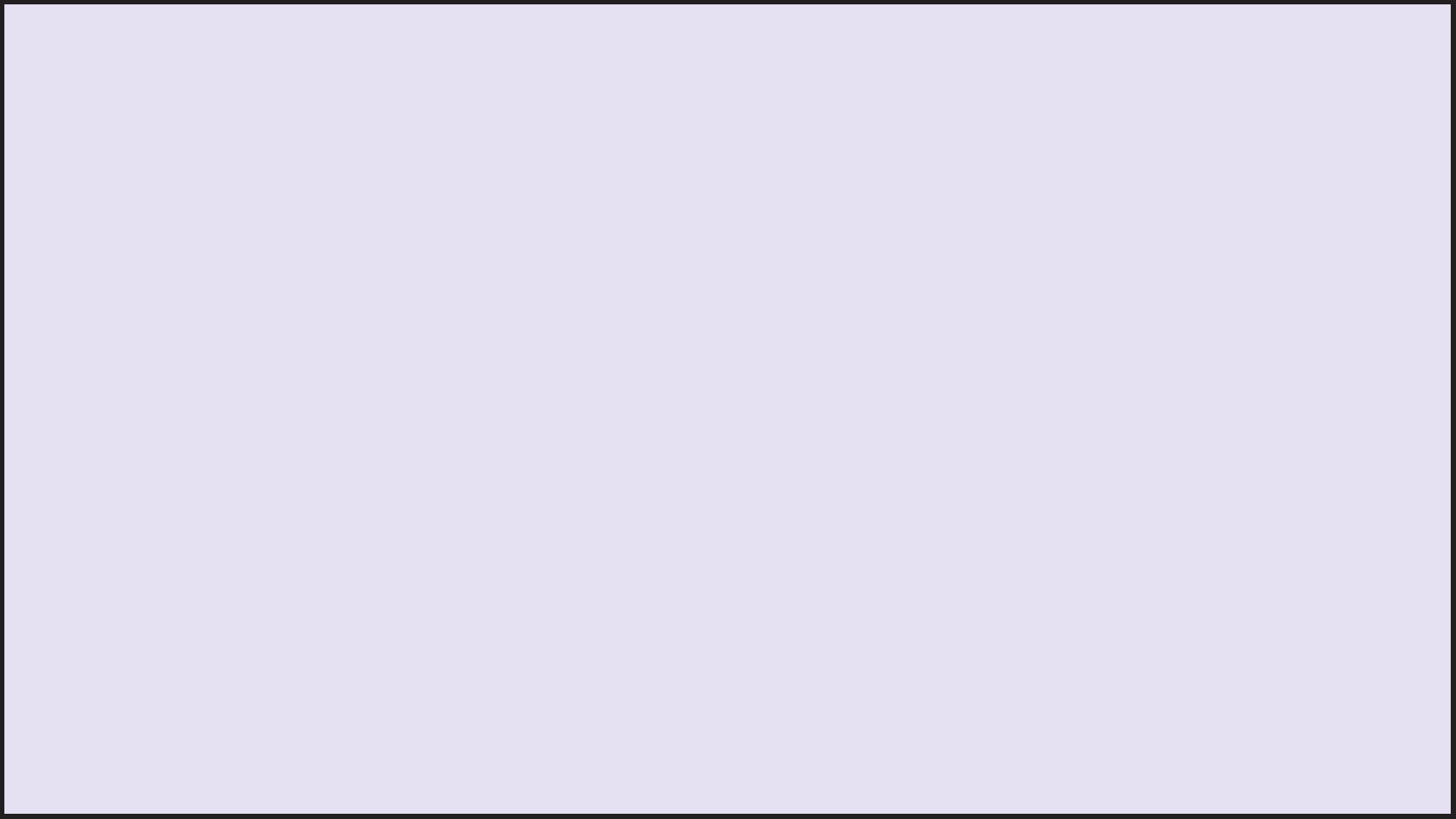} : 26\%--50\%, \includegraphics[width=8pt]{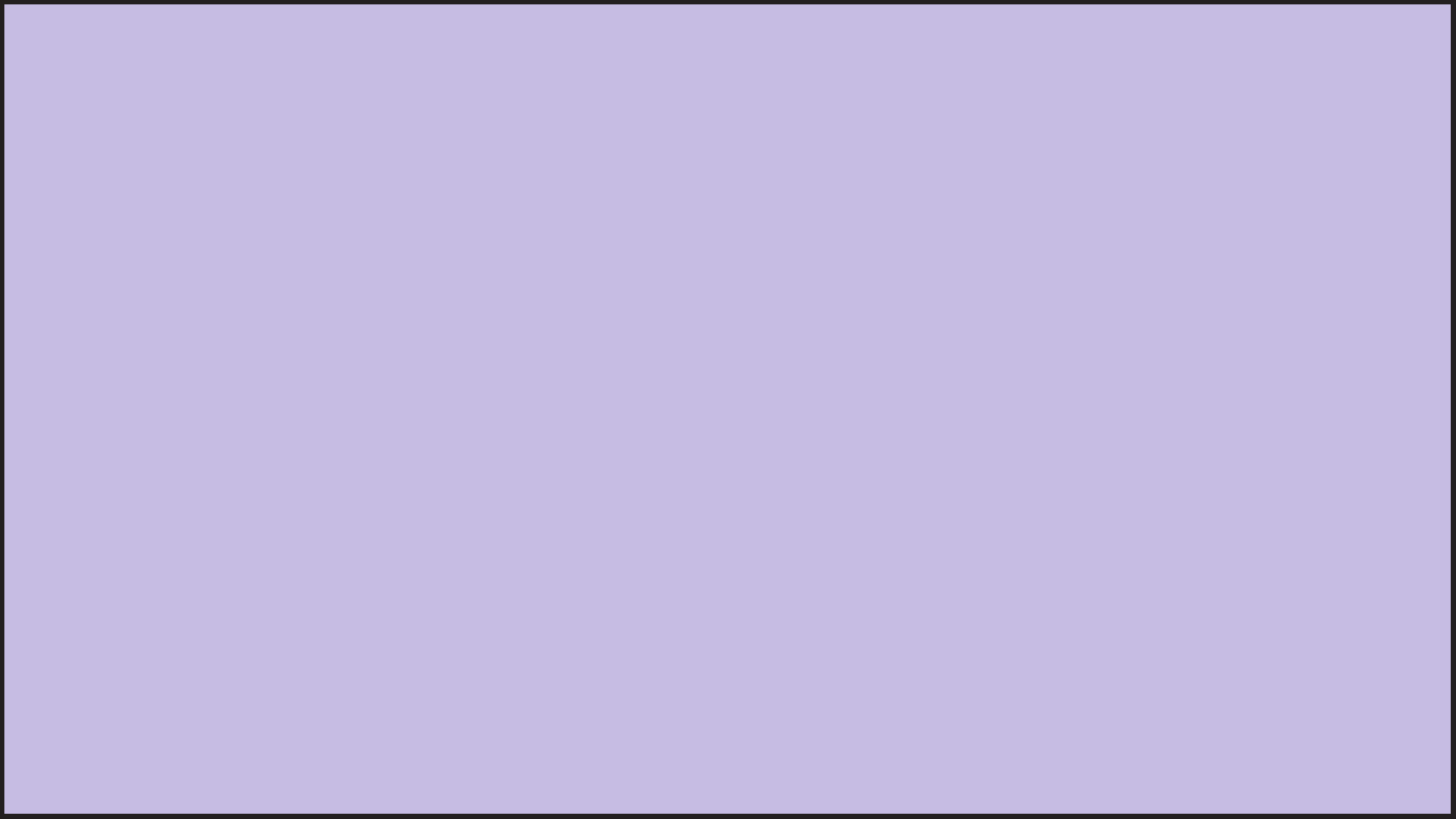} : 51\%--75\%, \includegraphics[width=8pt]{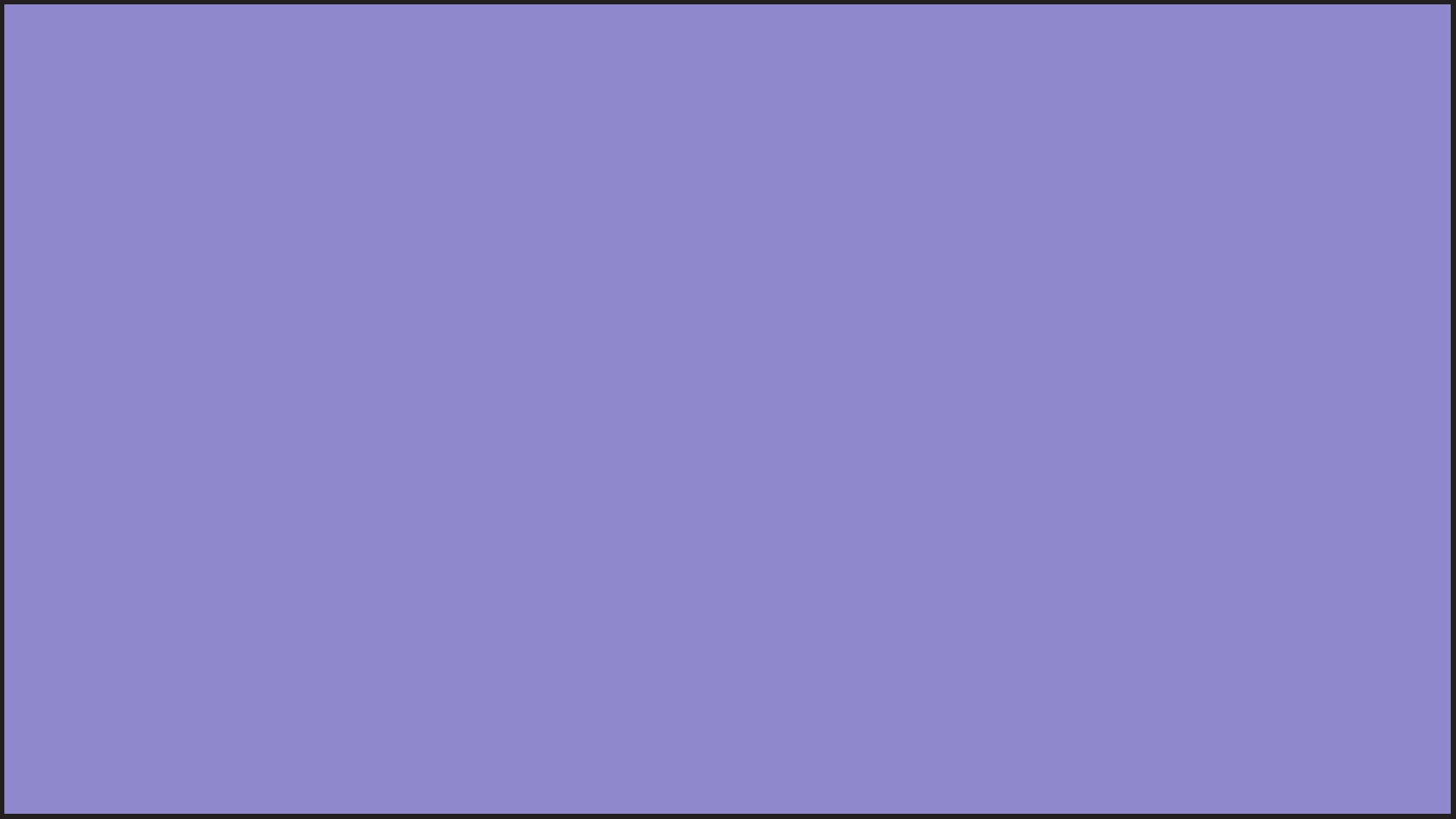} : 76\%--100\%.}
\vspace{-0.05in}
% \resizebox{\linewidth}{!}{
\begin{tabular}{cccccccccc}
%{\hsize}{@{}@{\extracolsep{\fill}}cccccccccccccc@{}}
\hline
\textbf{Target Model} 
& \textbf{Adv-Patch$\square$} 
& \textbf{Adv-Cloak$\square$}  
& \textbf{Adv-Texture$\square$} 
& \textbf{Adv-Mask$\square$} 
& \textbf{Adv-Sticker$\blacksquare$}
& \textbf{RL-Patch$\blacksquare$} 
& \textbf{HardBeat$\blacksquare$} 
& \textbf{RS-Patch$\blacksquare$} 
& \textbf{\sysname$\blacksquare$}  \\
\hline\hline

\multicolumn{1}{c|}{Yolov5-Face} 
& \cellcolor{mycolor4}95.26  
& \cellcolor{mycolor3}63.39   
& \cellcolor{mycolor3}65.57   
& \cellcolor{mycolor3}62.80  
& \cellcolor{mycolor1}25.27
& \cellcolor{mycolor3}61.09
& \cellcolor{mycolor4}83.82  
& \cellcolor{mycolor4}\textbf{100.00} 
& \cellcolor{mycolor4} 92.72  \\
%\hline
\multicolumn{1}{c|}{Yolov8-Face} & \cellcolor{mycolor4}92.17 & \cellcolor{mycolor4}89.82   & \cellcolor{mycolor4}87.82  
& \cellcolor{mycolor4}95.80  
& \cellcolor{mycolor4}84.18 
& \cellcolor{mycolor4}\textbf{100.00}  
& \cellcolor{mycolor4}\textbf{100.00}
& \cellcolor{mycolor4}\textbf{100.00}
& \cellcolor{mycolor4}\textbf{100.00}  \\
%\hline 
\multicolumn{1}{c|}{RetinaFace} & \cellcolor{mycolor3}58.47 & \cellcolor{mycolor2}48.73   & \cellcolor{mycolor3}55.27  
& \cellcolor{mycolor3}56.18    
& \cellcolor{mycolor3}61.82
& \cellcolor{mycolor4}90.18 
& \cellcolor{mycolor4}\textbf{96.00}  
& \cellcolor{mycolor4}91.09 
& \cellcolor{mycolor4}\textbf{99.09}  \\
%\hline

\multicolumn{1}{c|}{SCRFD} & \cellcolor{mycolor4}82.36 & \cellcolor{mycolor4}89.09  & \cellcolor{mycolor4}83.45  
& \cellcolor{mycolor4}90.18   
& \cellcolor{mycolor2}32.18
& \cellcolor{mycolor4}55.09
& \cellcolor{mycolor4}\textbf{100.00} & \cellcolor{mycolor4}\textbf{100.00}
& \cellcolor{mycolor4}\textbf{100.00} \\
%\hline
\multicolumn{1}{c|}{MTCNN} & \cellcolor{mycolor1}10.18 & \cellcolor{mycolor1}8.91  & \cellcolor{mycolor1}10.36  
& \cellcolor{mycolor1}13.09  
& \cellcolor{mycolor1}1.09 
& \cellcolor{mycolor2}28.00
& \cellcolor{mycolor3}67.09 & \cellcolor{mycolor4}86.91 
& \cellcolor{mycolor4}\textbf{92.91} \\
%\hline
\multicolumn{1}{c|}{ULFD} & \cellcolor{mycolor1}18.73& \cellcolor{mycolor1}4.55   & \cellcolor{mycolor1}25.27  
& \cellcolor{mycolor2}36.18    
& \cellcolor{mycolor2}36.00
& \cellcolor{mycolor2}41.09
& \cellcolor{mycolor3}71.09 & \cellcolor{mycolor3}74.91
&\cellcolor{mycolor4}\textbf{82.55} \\
%\hline
\multicolumn{1}{c|}{MogFace} & \cellcolor{mycolor2}47.09& \cellcolor{mycolor3}55.64   & \cellcolor{mycolor2}48.55 
& \cellcolor{mycolor3}70.18  
& \cellcolor{mycolor2}42.00
& \cellcolor{mycolor3}60.00 
& \cellcolor{mycolor3}64.55  & \cellcolor{mycolor4}85.27 
& \cellcolor{mycolor4}\textbf{85.71}\\
%\hline
\multicolumn{1}{c|}{Dlib} & \cellcolor{mycolor4}94.18& \cellcolor{mycolor1}13.09   & \cellcolor{mycolor3}55.09 
& \cellcolor{mycolor4}93.82   
& \cellcolor{mycolor1}14.55
& \cellcolor{mycolor4}97.09
& \cellcolor{mycolor4}\textbf{100.00}& \cellcolor{mycolor4}\textbf{100.00}
& \cellcolor{mycolor4}\textbf{100.00}\\
%\hline
\multicolumn{1}{c|}{OpenCV} & \cellcolor{mycolor3}79.78& \cellcolor{mycolor4}83.82   & \cellcolor{mycolor2}45.09 
& \cellcolor{mycolor4}81.09   
& \cellcolor{mycolor3}52.73
& \cellcolor{mycolor4}80.55
& \cellcolor{mycolor4}\textbf{100.00}  & \cellcolor{mycolor4}\textbf{100.00}
& \cellcolor{mycolor4}98.36\\
\hline
\end{tabular}
% }
% \caption{ASRs (\%) results of \sysname's digital attack in the visual images. Specifically, the white-box ($\square$) attack methods are trained on the Yolov5-Face model and tested on other models, and the black-box ($\blacksquare$) attack methods are directly tested on all models. ASR color bricks in tables: \includegraphics[width=8pt]{pictures/experiment/c1.pdf} : 0--25\%, \includegraphics[width=8pt]{pictures/experiment/c2.pdf} : 26\%--50\%, \includegraphics[width=8pt]{pictures/experiment/c3.pdf} : 51\%--75\%, \includegraphics[width=8pt]{pictures/experiment/c4.pdf} : 76\%--100\%.}
% \vspace{-0.25in}
\label{tab:digital attack visual domain}
\vspace{-0.15in}
\end{table*}